\documentclass[twocolumn,times]{aastex63}
 \makeatletter       
\let\@dates\relax
\makeatother
\usepackage[utf8]{inputenc}
\usepackage{amsmath, amsthm, amssymb}
\usepackage{bm}
\usepackage{graphicx}
\usepackage{xcolor}
\usepackage{xfrac}
\usepackage[normalem]{ulem}     % for strikethrough of text using \sout

\newcommand{\Lagr}{\mathcal{L}}
\let\vec\mathbf
\usepackage{natbib}
\usepackage{float}

\newcommand{\cnsq}{C_{\rm n}^2}

\begin{document}

\title{Electron Density Structure of the Local Galactic Disk}
\author[0000-0002-4941-5333]{Stella Koch Ocker}
\author[0000-0002-4049-1882]{James M. Cordes}
\author[0000-0002-2878-1502]{Shami Chatterjee}
\affiliation{Department of Astronomy and Cornell Center for Astrophysics and Planetary Science, Cornell University, Ithaca, New York, 14850, USA}
\correspondingauthor{Stella Koch Ocker}
\email{sko36@cornell.edu}

\shorttitle{Local Galactic Disk Structure}
\shortauthors{Ocker, S.K., Cordes, J.M., \& Chatterjee, S.}
\keywords{Galaxy: local interstellar matter --- Galaxy: structure --- pulsars: general}

\received{2020 April 27}
\revised{2020 May 29}
\accepted{2020 June 1}
\submitjournal{The Astrophysical Journal}

\begin{abstract}
 Pulsar dispersion measures (DMs) have been used to model the electron density of the interstellar medium (ISM) \replaced{above and below the Galactic plane}{in the Galactic disk} as a plane-parallel medium, despite significant scatter in the DM-distance distribution and strong evidence for inhomogeneities in the ISM. We use a sample of pulsars with independent distance measurements to evaluate a model of the local ISM \added{in the thick disk of the Galaxy} that incorporates turbulent fluctuations, clumps, and voids in the electron density. The latter two components are required because $\sim 1/3$ of the lines of sight are discrepant from a strictly plane parallel model. A likelihood analysis for smooth components of the model yields a scale height $z_0=1.57^{+0.15}_{-0.14}$ kpc and a mid-plane density $n_0=0.015 \pm 0.001$ cm$^{-3}$. The scatter in the DM-distance distribution is dominated by clumps and voids but receives significant contributions from a broad spectrum of density fluctuations, such as a Kolmogorov spectrum. The model is used to identify lines of sight with outlier values of DM. Three of these pulsars, J1614$-$2230, J1623$-$0908, and J1643$-$1224, lie behind known HII regions, and the electron density model is combined with H$\alpha$ intensity data to constrain the filling factors and other substructure properties of the HII regions (Sh 2-7 and Sh 2-27). Several pulsars also exhibit enhanced DM fluctuations that are likely caused by their lines of sight intersecting the superbubble GSH 238+00+09.
\end{abstract}

\section{Introduction}
An accurate Galactic electron density model is crucial for estimating pulsar distances when independent distance measurements are unavailable, in addition to estimating effects of the ISM on pulsar timing for gravitational wave detection. The main \replaced{tool of}{input to} such a model is pulsar dispersion. Pulsar dispersion measures (DMs) combined with parallax distances can be used to estimate the mean electron density $n_e$ of the ISM along a given line of sight (LoS) through the relation DM = $\int_0^D n_e {\rm{d}}l$ pc cm$^{-3}$, where $D$ is the distance to the pulsar. Extant models of Galactic electron density divide the Galactic ISM into thin and thick disk components, with more complex models like NE2001 \citep{2002astro.ph..7156C} and YMW16 \citep{2017ApJ...835...29Y} including features like spiral arms and the Local Bubble. 

\indent Understanding the distribution of free electrons in the Galactic ISM is also needed for the interpretation of angular scattering, scintillations, and plasma dispersion of extragalactic sources, including intraday variable sources and fast radio bursts (FRBs) \citep[e.g.,][]{2017ApJ...845...90V, 2019A&ARv..27....4P}.  FRBs, in particular, receive contributions to DM from the Milky Way, the intergalactic medium (IGM), and their host galaxy.  Separation of measured DMs into these three components requires an accurate foreground model for the Galaxy whether or not a redshift has been measured for the host galaxy. Similarly,  modeling scattering and scintillation from foreground plasma is needed to interpret the flux density distributions of FRBs, particularly those that repeat. 

\indent NE2001 is the most widely used Galactic electron density model, but has been shown to inaccurately predict some DMs at high Galactic latitudes \citep{2008PASA...25..184G, 2018ApJ...864...26J}. Pulsars at high latitudes sample the thick, warm ionized component of the ISM as well as other components in the thin disk or local ISM.  Using DMs and independent distance measurements, numerous studies \citep{1977ApJ...215..885T, 1982JApA....3..399V, 1989ApJ...339L..29R, 1992AJ....104.1465N, 2008PASA...25..184G, 2009ApJ...702.1472S, 2012MNRAS.427..664S} have shown that the electron density at these latitudes roughly follows a plane-parallel model of the form \replaced{$n_e = n_0 z_0e^{-|z|/z_0}$}{$n_e = n_0 e^{-|z|/z_0}$}, where $n_0$ is the mean electron density at mid-plane, $z_0$ is the electron density scale height, and $z$ is the vertical distance \replaced{above or below}{from} the Galactic plane. Since the plane-parallel model has a sharp cusp at $z=0$, the NE2001 and \cite{1993ApJ...411..674T} (TC93) Galactic models use a more physical $z$ dependence of the form sech$^2$($|z|/z_0$) \added{for the thick disk component of the Galaxy}. \replaced{Both the plane-parallel and hyperbolic secant models have perpendicular DMs that asymptote at $n_0z_0$, but the scale heights of these models are not equivalent, as the integrated column density of the hyperbolic secant model asymptotes at lower $|z|$.}{In both the plane-parallel and hyperbolic secant models, the integrated column density of the entire Galactic disk has an asymptotic value of $n_0 z_0$, but the scale heights of these models are not equivalent, as the integrated column density of the hyperbolic secant model reaches $n_0 z_0$ at lower $|z|$. Similar to previous studies, our focus is on the $|z|$ dependence of electron density, and it is implicit that the plane-parallel component is restricted in Galactocentric radius.}

\indent Over the past decade, empirical studies have \replaced{contended the}{estimated} the scale height of the plane-parallel model and found it ranges from 1410 pc \citep{2009ApJ...702.1472S} to 1830 pc \citep{2008PASA...25..184G}, and they have consistently argued that NE2001's scale height of 1 kpc is therefore too low. We note that despite these differences in scale height, the models generally agree with NE2001 on the integrated column density through the entire disk ($n_0 z_0 \approx 23$ pc cm$^{-3}$). The most recently created Galactic model, YMW16, also departs from empirical scale height studies, since it uses a scale height of 1675 pc for the extended thick disk component, but has a significantly lower total integrated column density, $n_0 z_0 = 18.9 \pm 0.9$ pc cm$^{-3}$.

\indent All previous studies were based on limited pulsar samples and did not meaningfully constrain inhomogeneities along the lines of sight. This study aims to not only update the thick disk modeling to include new pulsar distance measurements, but to also account for the effects of turbulence and discrete structures in the relatively local ISM. 

\indent Turbulence in interstellar plasma is probed by a variety of pulsar observations, from temporal DM variations to pulse broadening times and scintillation. Numerous studies over the past few decades have shown that pulsar observations imply an electron density power spectrum broadly consistent with a Kolmogorov power law up to an outer spatial scale of about $10^{18}$ meters \citep[e.g.,][]{1995ApJ...443..209A, 2015ApJ...804...23K}. Most recently, \cite{2019NatAs...3..154L} used Voyager mission detections of plasma oscillations just outside the heliopause to demonstrate that the Kolmogorov power law can extend down to even smaller scales than those probed with pulsars. Previous estimates of the scale height and mid-plane density of the Galactic disk did not include the effect of turbulence in their models \citep[e.g.,][]{2008PASA...25..184G}, or they incorporated a DM variance that was not based on a physical model of interstellar turbulence \citep[e.g.,][]{2009ApJ...702.1472S}. 

\indent Using the most updated sample of pulsar distance measurements made with parallaxes and globular cluster associations, we investigate a model for the electron density of the Galactic disk consisting of a plane-parallel medium with density fluctuations due to Kolmogorov turbulence. We identify lines of sight with DMs that depart significantly from the model owing to the presence of foreground HII regions. Constraints on the internal properties of these regions are made using the electron density model. Section 2 describes the model's theoretical framework, Section 3 summarizes the sample of pulsar distance measurements, and Sections 4 and 5 describe the results of fitting the model to the pulsars' distances and using deviations from the model to identify and characterize discrete structures in the ISM. Conclusions and plans for future work are given in Section 6.

\indent We use the following notation throughout the paper: the electron density model consisting of a plane-parallel medium with Kolmogorov density fluctuations is referred to as the PPK model. \added{The integrated column density predicted by the model is referred to as $N_e$ to distinguish it from observed pulsar DMs, and the following shorthand is used for the perpendicular component of observed DM and the corresponding model estimate: DM$_\perp$ = DMsin($|b$) and $N_{e\perp}$.} When mentioned, dispersion measure (DM), scattering measure (SM), and emission measure (EM) are in their standard units of pc cm$^{-3}$, kpc m$^{-20/3}$, and pc cm$^{-6}$, respectively. 

\section{The Electron Density Model}
\added{Full} Galactic electron density models like TC93, NE2001, and YMW16 adopt a multi-component structure in which a homogeneous medium serves as a baseline to which more realistic complexities are added. \added{To model the thick disk,} we adopt the same approach by adding Kolmogorov density fluctuations to the plane-parallel medium. While the wavenumber spectrum along each LoS in our sample does not exactly follow a Kolmogorov power law, in most cases it is a good benchmark model of the spectrum. In the electron density model, the plane-parallel medium predicts the mean electron density (and hence mean DM), while the Kolmogorov turbulence predicts the variance in that density/DM. The mean and variance construct a likelihood function that describes the probability of observing a certain DM given a distance.
\subsection{Plane-Parallel Spatial Dependence}
We use a plane-parallel model for the mean electron density with a $z$ dependence
\begin{equation}\label{eq:smooth}
    \bar{n}_{e,0}(z) = n_0e^{-|z|/z_0},
\end{equation}
where $n_0$ is the mid-plane electron density, $|z| = D$sin$(|b|)$ is the perpendicular distance from the Galactic plane, $b$ is the pulsar latitude, and $z_0$ is the scale height. \replaced{The perpendicular DM (DM$_{\perp}=$ DMsin$(|b|)$)}{The integrated column density perpendicular to the plane ($N_{e\perp} = \langle N_e \rangle {\rm sin}(|b|)$}) is then
%\begin{equation}\label{eq:ppmod}
%    {\rm{DM}_\perp}(z) = n_0 z_0 (1 - e^{-|z|/z_0}).
%\end{equation}
\begin{equation}\label{eq:ppmod}
    {N_{e\perp}}(z) = n_0 z_0 (1 - e^{-|z|/z_0}).
\end{equation}
\added{We refer to the integrated column density predicted by the model as $N_e$ to distinguish it from the observed pulsar DMs used to fit the model.} The maximum \replaced{DM$_\perp$}{$N_{e\perp}$} contributed by the entire Galactic disk is \replaced{DM$_{\perp,\infty}$}{${N}_{e \perp}(\infty) = n_0 z_0$}. This deterministic model is augmented by adding Kolmogorov fluctuations and clumps and voids to the electron density. 

\subsection{Kolmogorov Density Fluctuations}
We include electron density fluctuations with an isotropic Kolmogorov spectrum to give a total density of the form
\begin{equation}
    n_e(\vec{x}) = \bar{n}_{e,0}(z) + n_{k}(\vec{x}).
\end{equation}
The density fluctuations are a zero mean process with a second moment that is related to the Kolmogorov spectrum:
\begin{equation}
    \langle n_k^2 \rangle = \int {\rm{d}}^3q P_{n_k}(s,q),
\end{equation}
where the spectrum is the product of a coefficient $\cnsq(s)$ that varies slowly along a line of sight and a power-law in wavenumber $q$:
\begin{equation}
    P_{n_k}(s,q) = \cnsq(s)q^{-11/3},\hspace{0.1in} \frac{2\pi}{l_0} \leq q \leq \frac{2\pi}{l_1}
\label{eq:spectrum}
\end{equation}
\citep[e.g.,][]{1975ApJ...202..439L, 1985ApJ...288..221C, 1990ARA&A..28..561R, 2002astro.ph..7156C}. The spectrum extends from the outer scale $l_0$ to the inner scale $l_1$. Location along a line of sight of length $D$ is given by $0 \le s \le D$.  In practice the outer scale is much larger (on the order of kiloparsecs) than the inner scale, so we assume $q_0 \sim 10^{-18}$ m$^{-1}$ and $q_1 \rightarrow \infty$ \citep[e.g.,][]{1995ApJ...443..209A}.\\
\indent A likelihood function is constructed to fit the model to the data. \deleted{We refer to the DM predicted by this model as $N_e$ to distinguish the predicted dispersion from the measured dispersion when fitting the model.} The likelihood function $\Lagr$ for the model parameters $\Theta$ given observations of DM and distance ($D$) is
\begin{equation}\label{eq:ll}
\begin{split}
    \Lagr(\Theta) &= \prod_j \int f_{D_j}(D_j) {\rm{d}}(D_j) \bigg\{[2\pi \sigma_{N_e}^2(\hat{n}_j,D_j)]^{-1/2}\\
    &\times e^{-[{\rm{DM}}_j - \langle N_e(\hat{n}_j,D_j) \rangle]^2/2\sigma_{N_e}^2(\hat{n}_j,D_j)}\bigg\},
\end{split}
\end{equation}
where the distances $D_j$ are derived from parallaxes and globular cluster associations whose errors are described well by a Gaussian probability density function (PDF), which yields a PDF for the distance $f_{D_j}(D_j)$. The DM variance of this model depends on the scattering measure (SM = $\int_0^D {\rm{d}}s \cnsq$):
\begin{equation}
    \sigma_{N_e}^2 = (3/5){\rm{SM}}q_0^{-5/3}.
\label{eq:varN}
\end{equation}
A derivation of the likelihood function is given in Appendix~\ref{app:theory}.

\section{Pulsar Distances}
There are over 140 pulsar parallax measurements in the published literature\footnote{see the current list at \url{www.astro.cornell.edu/research/parallax/}, last updated on March 2, 2020.}, obtained through Very Long Baseline Interferometry (VLBI), optical techniques (e.g. with the Hubble Space Telescope or Gaia DR2), and pulsar timing. We limit the sample to pulsars with fractional parallax uncertainties less than $25\%$, and to avoid the complexities of the inner Galactic disk and spiral arms, we exclude pulsars with latitudes $|b|<20^{\circ}$, leaving the 37 pulsars shown in Table~\ref{tab:pulsarsample}. For pulsars with multiple parallax measurements, we use the parallaxes with the smallest fractional uncertainties.
The pulsar parallaxes $\pi$ are converted to distances from the Galactic plane $|z|$ according to $|z| = (1/\pi){\rm{sin}}(|b|)$,
and the errors propagated accordingly.\\
\indent Pulsars have also been localized to 28 globular clusters\footnote{see the current list at \url{www.naic.edu/~pfreire/GCpsr.html}, last updated on February 25, 2020.}. For each cluster with more than one pulsar we calculate the average DM and take the uncertainty in the associated distance to be $15\%$ \citep[][]{2012MNRAS.427..664S}. In each case the $1\sigma$ error in the DM is much less than the uncertainty in the distance. The globular cluster sample is also limited to $|b|>20^{\circ}$, leaving the nine clusters shown in Table~\ref{tab:pulsarsample}. About 30 pulsars have also been discovered in the Magellanic Clouds, but are not included here due to the uncertainty in the Clouds' contributions to their DMs \citep{1992AJ....104.1465N, 2013MNRAS.433..138R}.

\begin{deluxetable*}{l l R R R R R R R}
\tablecaption{Distance Measurements Beyond $\pm 20^{\circ}$ Galactic Latitude
\label{tab:pulsarsample}}
\tabletypesize{\scriptsize}
\tablewidth{0pt}
\tablehead{\multicolumn{2}{c}{Pulsar} & \colhead{$l$} & \colhead{$b$} & \colhead{DM} & \colhead{DM$_\perp$} & \colhead{$\pi$} & \colhead{$D$} & \colhead{$|z|$} \\
\colhead{J-Name} & \colhead{B-Name} & \multicolumn{2}{c}{(Degrees)} & \multicolumn{2}{c}{(pc cm$^{-3}$)} & \colhead{(mas)} & \colhead{(kpc)} & \colhead{(kpc)}}
\tablecolumns{9}
\startdata
 J0034$-$0721 & B0031$-$07 & 110.420& -69.815 & 10.922 & 10.25 & 0.93^{+0.08}_{-0.07} & 1.075^{+0.087}_{-0.085} & 
 1.009^{+0.081}_{-0.085}\\
 J0437$-$4715$^\dagger$ & & 253.394& -41.963 & 2.645 & 1.77 & 6.396^{+0.054}_{-0.054}& 0.156^{+0.001}_{-0.001} & 0.104^{+0.001}_{-0.001}\\ 
 J0751+1807$^\dagger$ & & 202.730& 21.086 & 30.246 & 10.88 &0.66^{+0.15}_{-0.15} & 1.51^{+0.45}_{-0.28} & 0.55^{+0.16}_{-0.10} \\
 J0814+7429$^\dagger$ & B0809+74 & 139.998& 31.618 & 5.751 & 3.01 & 2.31^{+0.04}_{-0.04}& 0.433^{+0.007}_{-0.007} & 0.227 ^{+0.004}_{-0.004}\\
 J0826+2637$^\dagger$ & B0823+26 & 196.964& 31.743 & 19.476 & 10.25 & 2.010^{+0.013}_{-0.009}& 0.497^{+0.002}_{-0.003} & 0.262^{+0.001}_{-0.002} \\
 J0922+0638$^\dagger$ & B0919+06 & 225.420& 36.392 & 27.2986 & 16.20 & 0.83^{+0.13}_{-0.13} & 1.20^{+0.22}_{-0.16} & 0.71^{+0.13}_{-0.10} \\
 J0953+0755$^\dagger$ & B0950+08 & 228.908& 43.697 & 2.969 & 2.05 & 3.82^{+0.07}_{-0.07}& 0.261^{+0.005}_{-0.005}& 0.181^{+0.003}_{-0.003}\\
 J1022+1001$^\dagger$ & & 231.795& 51.101 & 10.252 & 7.98 & 1.387^{+0.041}_{-0.028}& 0.72^{+0.015}_{-0.020} & 0.56^{+0.01}_{-0.02}\\
 J1023+0038$^\dagger$ & & 243.490& 45.782 & 14.325 & 10.27 & 0.731^{+0.022}_{-0.022}& 1.36^{+0.04}_{-0.04} & 0.98^{+0.03}_{-0.03}\\
 J1024$-$0719$^\dagger$ & & 251.702& 40.515 & 6.477 & 4.21 & 0.89^{+0.14}_{-0.14} & 1.12^{+0.21}_{-0.15} & 0.73^{+0.14}_{-0.10}\\
 J1136+1551$^\dagger$ & B1133+16 & 241.895& 69.196 & 4.8407 & 4.53 & 2.687^{+0.018}_{-0.016}& 0.372^{+0.002}_{-0.002} & 0.348^{+0.002}_{-0.002}\\
 J1239+2453$^\dagger$ & B1237+25 & 252.450& 86.541 & 9.2515 & 9.23 & 1.16^{+0.08}_{-0.08}& 0.86^{+0.06}_{-0.06} & 0.86^{+0.06}_{-0.06}\\
 J1321+8323 & B1322+83 & 121.887& 33.672 & 13.316 & 7.38 & 0.97^{+0.04}_{-0.14}& 1.03^{+0.17}_{-0.04} & 0.57^{+0.10}_{-0.02}\\
 J1455$-$3330 & & 330.722& 22.562 & 13.5698 & 5.21 & 0.99^{+0.22}_{-0.22}& 1.01^{+0.28}_{-0.18} & 0.39^{+0.11}_{-0.07}\\
 J1509+5531$^\dagger$ & B1508+55 & 91.325& 52.287 & 19.619 & 15.52 & 0.47^{+0.03}_{-0.03}& 2.12^{+0.14}_{-0.13} & 1.68^{+0.11}_{-0.10}\\
 J1532+2745 & B1530+27 & 43.481& 54.495 & 14.691 & 11.96 & 0.62^{+0.03}_{-0.10}& 1.61^{+0.31}_{-0.07} & 1.31 ^{+0.25}_{-0.06}\\
 J1537+1155$^\dagger$ & B1534+12 & 19.848& 48.341 & 11.6194 & 8.68 & 0.96^{+0.01}_{-0.01}& 1.04^{+0.01}_{-0.01} & 0.778^{+0.008}_{-0.008}\\
 J1543+0929 & B1541+09 & 17.811& 45.775  & 34.9758 & 25.06 & 0.13^{+0.02}_{-0.02}& 7.69^{+1.39}_{-1.02} & 5.51^{+1.00}_{-0.73}\\
 J1543$-$0620 & B1540$-$06 & 0.566& 36.608 & 18.3774 & 10.96  & 0.322^{+0.028}_{-0.045}& 3.10^{+0.50}_{-0.25} & 1.85^{+0.30}_{-0.15}\\
 J1607$-$0032 & B1604$-$00 & 10.715& 35.466 & 10.682 & 6.20 & 0.910^{+0.029}_{-0.046}& 1.09^{+0.06}_{-0.03} & 0.64^{+0.03}_{-0.02}\\
 J1614$-$2230$^\dagger$ & & 352.636& 20.192 & 34.918 & 12.05 & 1.3^{+0.09}_{-0.09} & 0.77^{+0.06}_{-0.05} & 0.27^{+0.02}_{-0.02}\\
 J1623$-$0908 & & 5.297& 27.178 & 68.183 & 31.14 & 0.59^{+0.1}_{-0.1} & 1.69^{+0.34}_{-0.24} & 0.77^{+0.15}_{-0.11}\\
 J1643$-$1224$^\dagger$ & & 5.669& 21.218 & 62.4143 & 22.59 & 2.2^{+0.4}_{-0.4} & 0.45^{+0.10}_{-0.07} & 0.16^{+0.04}_{-0.03}\\
 J1645$-$0317$^\dagger$ & B1642–03 & 14.114& 26.062 & 35.7555 & 15.71 & 0.252^{+0.028}_{-0.019}& 3.97^{+0.32}_{-0.39} & 1.74^{+0.14}_{-0.17}\\ 
 J1713+0747$^\dagger$ & & 28.751& 25.223 & 15.917 & 6.78 & 0.94^{+0.05}_{-0.05}& 1.06^{+0.06}_{-0.05} & 0.45^{+0.03}_{-0.02}\\
 J1754+5201 & B1753+52 & 79.608& 29.628 & 35.0096 & 17.31 & 0.160^{+0.029}_{-0.022}& 6.25^{+0.99}_{-0.96} & 3.09^{+0.49}_{-0.47}\\
 J1840+5640 & B1839+56 & 86.078& 23.819 & 26.7716 & 10.81 & 0.657^{+0.065}_{-0.008} & 1.52^{+0.02}_{-0.13} & 0.61^{+0.008}_{-0.06}\\
 J2006$-$0807 & B2003$-$08 & 34.101& -20.304 & 32.39 & 11.24 & 0.424^{+0.010}_{-0.101} & 2.35^{+0.74}_{-0.05} & 0.82^{+0.25}_{-0.02}\\ 
 J2048$-$1616$^\dagger$ & B2045$-$16 & 30.514& -33.077 &  11.456 & 6.25 & 1.05^{+0.03}_{-0.02}& 0.95^{+0.02}_{-0.03} & 0.52^{+0.01}_{-0.01} \\
 J2124$-$3358$^\dagger$ & & 10.925& -45.438 &  4.60096 &  3.28 & 3.1^{+0.55}_{-0.55}& 0.32^{+0.07}_{-0.05} & 0.23^{+0.05}_{-0.03}\\
 J2129$-$5721$^\dagger$ & & 338.005& -43.560 & 31.851 & 21.95 & 0.424^{+0.088}_{-0.088}& 2.36^{+0.62}_{-0.40} & 1.63^{+0.43}_{-0.28}\\
 J2144$-$3933 & & 2.794& -49.466 & 3.35 & 2.55 & 6.05^{+0.56}_{-0.56}& 0.16^{+0.02}_{-0.01} & 0.13^{+0.01}_{-0.01}\\
 J2145$-$0750$^\dagger$ & & 47.777& -42.084 & 8.99 & 6.03 & 1.603^{+0.063}_{-0.009}& 0.624^{+0.003}_{-0.023} & 0.42^{+0.002}_{-0.02}\\
 J2222$-$0137 & & 62.018& -46.075 & 3.277 &  2.36 & 3.742^{+0.013}_{-0.016} & 0.2672^{+0.0011}_{-0.0009}  & 0.1925^{+0.0008}_{-0.0006}\\
 J2305+3100 & B2303+30 & 97.721& -26.657 & 49.5845 & 22.25 & 0.223^{+0.033}_{-0.028}& 4.48^{+0.64}_{-0.58} & 2.01^{+0.29}_{-0.26}\\
 J2317+2149 & B2315+21 & 95.831& -36.075 & 20.86959 & 12.29 & 0.51^{+0.06}_{-0.05}& 1.96^{+0.21}_{-0.20} & 1.15^{+0.13}_{-0.12}\\
 J2346$-$0609 & & 83.798& -64.015 & 22.504 & 20.23 & 0.275^{+0.021}_{-0.036}& 3.6^{+0.55}_{-0.26} & 3.27^{+0.49}_{-0.23}\\
 \hline \hline 
 \multicolumn{2}{c}{Globular Cluster} & \multicolumn{1}{c}{$l$} & \multicolumn{1}{c}{$b$} & \multicolumn{1}{c}{DM} & \multicolumn{1}{c}{DM$_\perp$} & \# \hspace{0.04in} {\rm pulsars} & \multicolumn{1}{c}{$D$} & \multicolumn{1}{c}{$|z|$} \\ 
 {NGC$\#$} & {a.k.a.} & \multicolumn{2}{c}{({Degrees})} & \multicolumn{2}{c}{(pc cm$^{-3}$)}  & {} & \multicolumn{1}{c}{(kpc)} & \multicolumn{1}{c}{(kpc)} \\ \hline
 NGC104 & 47Tuc & 305.90& -44.89 & 24.4 & 17.24 & 25 & 4.5 & 3.2 \\
 NGC1851 & & 244.51& -35.04 & 52.1 & 29.94 & 1 & 12.1 & 6.95 \\
 NGC5024 & M53 & 332.96& 79.76 & 24.0 & 23.62 & 1 & 17.8 & 17.5 \\
NGC5272 & M3 & 42.41& 78.71 & 26.4 & 25.86 & 3 & 10.4 & 10.2 \\
NGC5904  & M5 & 3.86& 46.8 & 29.5 & 21.49 & 5  &  7.5 & 5.5 \\
NGC6205 & M13& 59.01& 40.91 & 30.2 & 19.76 & 6 & 7.7 & 5.0 \\
NGC6752 & & 336.49& -25.63 & 33.4 & 14.43 & 5 & 4.0 & 1.7 \\
NGC7078 & M15 & 65.01& -27.31 & 66.9 & 30.68 & 8 & 10.3 & 4.73 \\
NGC7099 & M30 & 27.18& -46.83 & 25.1 & 18.29 & 2 & 8.0 & 5.8 \\ 
\enddata
\tablecomments{Pulsar parallaxes were obtained from Shami Chatterjee's website \url{www.astro.cornell.edu/research/parallax/} and the following references therein:
\cite{2001ApJ...550..287C},
\cite{2002ApJ...571..906B},
\cite{2008ApJ...685L..67D},
\cite{2009ApJ...698..250C}, \cite{2012ApJ...756L..25D}, \cite{2014ApJ...787...82F},
\cite{2016AA...587A.109G}, and
\cite{2019ApJ...875..100D}. Pulsars with multiple parallax measurements are indicated with a $\dagger$; in these cases, the parallax with smallest fractional error is shown. Globular cluster distances were obtained from Paulo Freire's website \url{www.naic.edu/~pfreire/GCpsr.html} and references therein. For each cluster with more than one pulsar we calculate the average DM and take the uncertainty in the associated distance to be $15\%$.}
\end{deluxetable*}

\section{Scale Height and Mid-Plane Density}

\indent The scale height $z_0$ and mid-plane density $n_0$ are constrained using measurements of perpendicular distance from the Galactic plane $|z|$ and DM$_\perp$ \added{= DMsin($|b|$)}. Previous scale height studies ignored pulsars with LoS through known HII regions to avoid sources with excess dispersion \citep[e.g.,][]{2008PASA...25..184G, 2012MNRAS.427..664S}. We identified pulsars with significant DM excess by fitting the model iteratively and finding those pulsars more than 3$\sigma$ away from the final best fit model (where $\sigma$ is the $68\%$ confidence interval). Hereafter, these pulsars are referred to as outliers. Seven of these outliers have excess DMs caused by identifiable discrete structures along the LoS and were excluded from the fit; these outliers are discussed in Section~\ref{sec:outliers}. The rest of the outliers are discussed in Appendix~\ref{app:outliers}.

\subsection{Contributions to the DM Variance}
\indent Kolmogorov electron density fluctuations (c.f. Eq.~\ref{eq:spectrum}) contribute  variance to DM according to Eq.~\ref{eq:varN}. We incorporate this contribution in the likelihood analysis, where the plane-parallel component yields the \replaced{mean DM}{mean integrated column density ($\langle N_e \rangle$)} and the Kolmogorov fluctuations yield the \replaced{DM variance}{variance $\sigma_{N_e}^2$}. The resulting likelihood function (c.f. Eq.~\ref{eq:ll}) is maximized when both the difference between \replaced{mean DM}{$\langle N_e \rangle$} and observed DM and the \replaced{DM variance}{variance $\sigma_{N_e}^2$} are minimal (compounded with the $|z|$ errors). 

\indent Density fluctuations most likely scale with the square of the local mean density (c.f. cloudlet model discussed in the next section), so we assume that $\cnsq$ follows a plane-parallel distribution with half the scale height of the electrons: $\cnsq = C_{\rm n,0}^2 e^{-2|z|/z_0}$, where $C_{\rm n,0}^2 = 10^{-3.5}$ m$^{-20/3}$. For most pulsars in the data set, this yields SMs on the order of $10^{-4}$ to $10^{-5}$ kpc m$^{-20/3}$ and $\sigma_{N_e}$ on the order of 1 to 3 pc cm$^{-3}$. However, for the highest $|z|$ objects in the sample ($|z| > 2$ kpc), the observed scatter in DM$_\perp$ is consistent with an empirical rms $\sigma_{\rm DM} \sim 10$ pc cm$^{-3}$, much larger than the rms predicted by the modeled turbulence. This rms DM is likely due to the longer path lengths for these LoS, which increase the probability of encountering clumps and voids. To account for the observed variance in the DMs of these high-$|z|$ pulsars, we add a fractional DM error of 0.15DM$_\perp$ for objects with $|z|>2$ kpc. We tested the likelihood optimization on a simulated sample of pulsars with DMs varying according to the PPK model and found agreement between simulated parameters and the maximum likelihood results with a $\bar{\chi}^2 = 0.03$. 

\begin{figure*}

\centering
\includegraphics[width=0.65\textwidth]{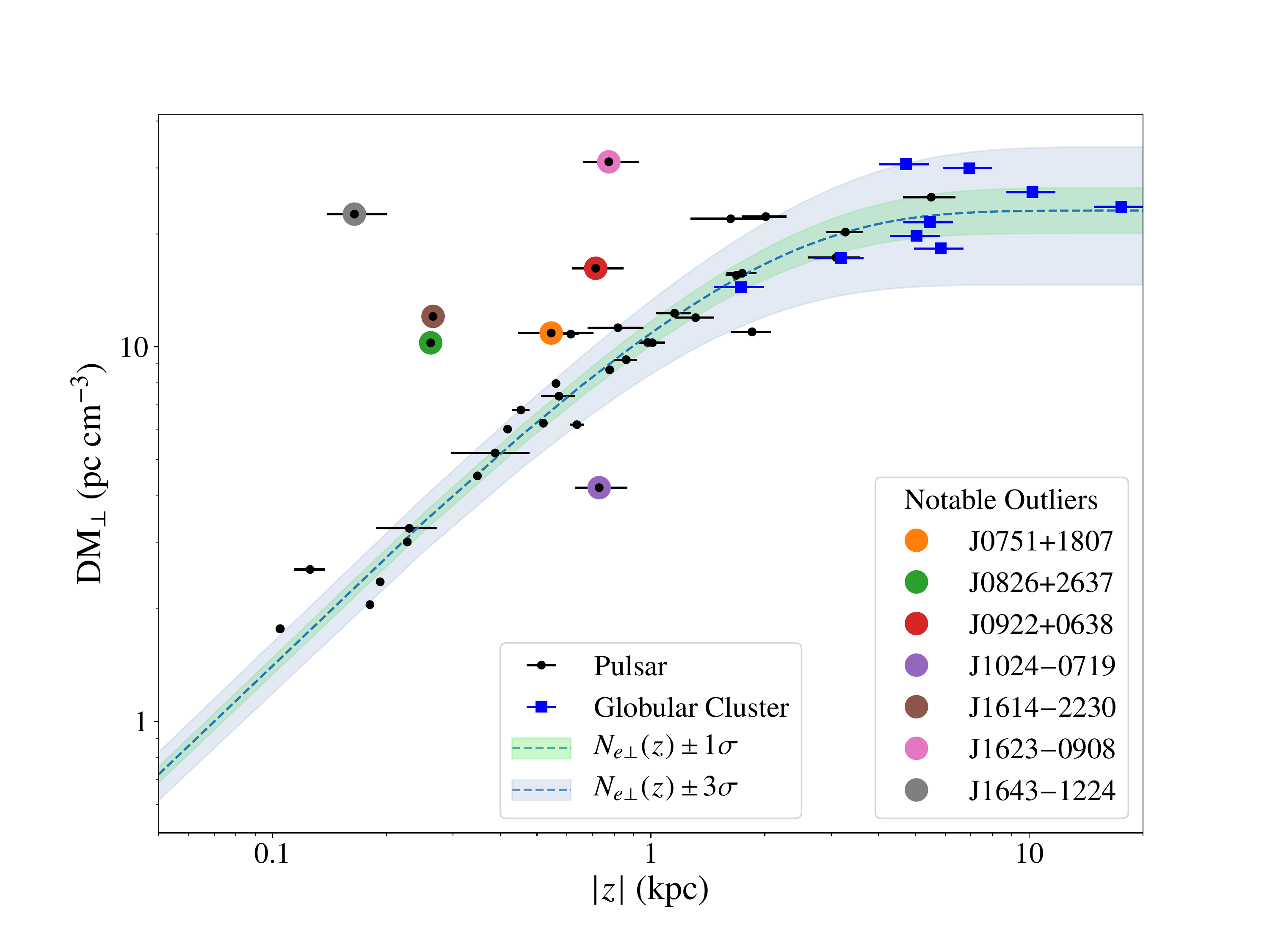}
\caption{The perpendicular component of DM vs. distance from the Galactic plane for objects with latitudes $|b|>20^{\circ}$. Blue squares indicate globular clusters harboring pulsars, while black points indicate pulsars whose distances were obtained either by timing, optical, or interferometric measurements of parallax. The dashed line indicates the plane-parallel component of the model (\added{$N_{e\perp}(z)$}), with $z_0 = 1.57^{+0.15}_{-0.14}$ kpc and $n_0 = 0.015 \pm 0.001$ cm$^{-3}$. The $\pm1\sigma$ and $\pm 3\sigma$ intervals are indicated by the shaded regions. Seven pulsars excluded from the fit are indicated by the colored circles; discrete structures intervene along these lines of sight.}
\label{fig:scaleheightfit}

\includegraphics[width=0.5\textwidth]{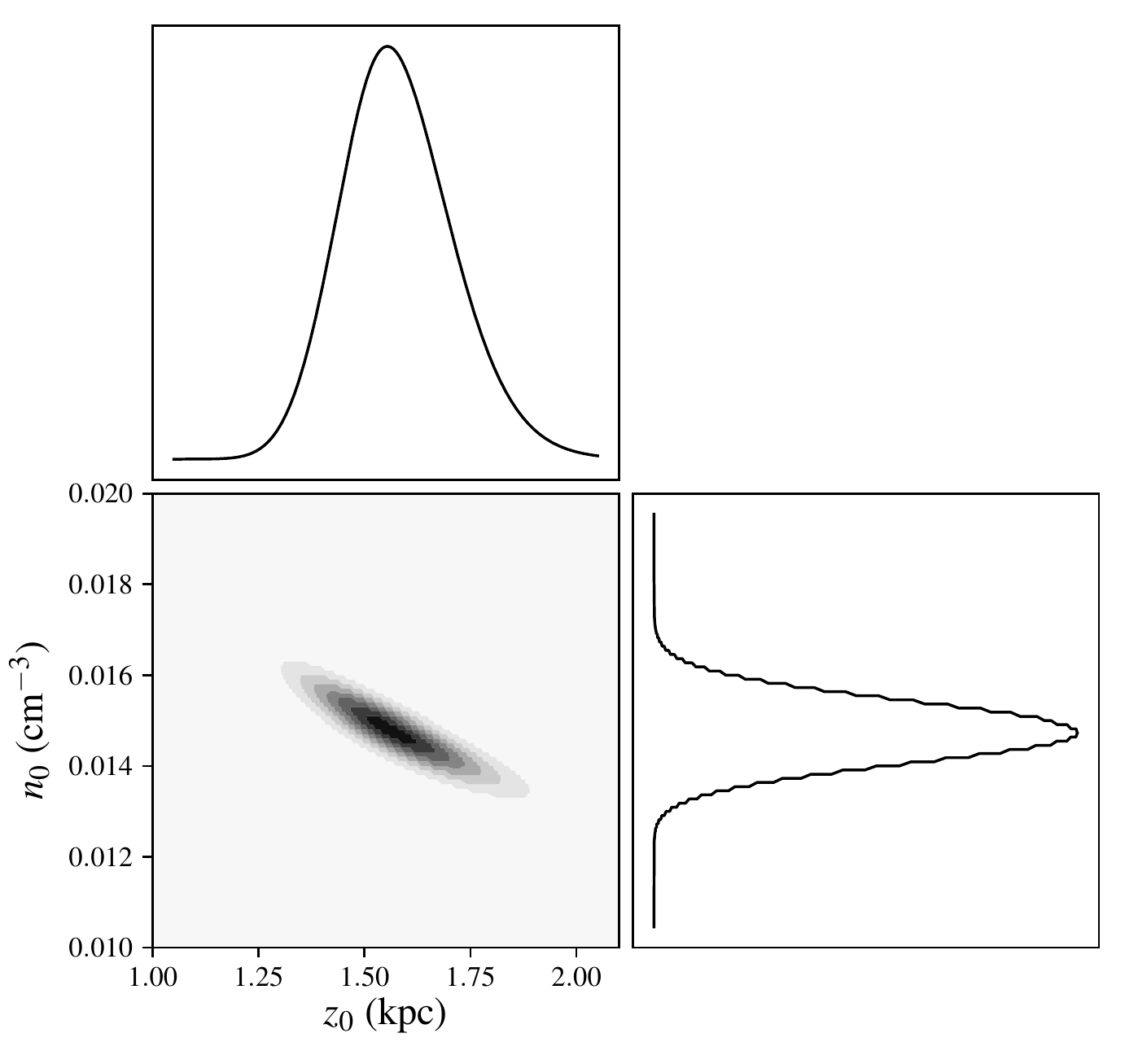}
    \caption{The 2D posterior and marginalized PDFs for the scale height $z_0$ and mid-plane density $n_0$, assuming flat priors on both parameters. The 50th, 16th, and 84th percentiles of the cumulative distribution function for each parameter were taken to be the best fit value, lower error, and upper error, respectively. The resulting constraints are $n_0 = 0.015 \pm 0.001$ cm$^{-3}$ and $z_0 =
    1.57^{+0.15}_{-0.14}$ kpc.}
    \label{fig:triangle}
    
\end{figure*}

\subsection{Fitting Results}

\indent The normalized posterior probability density function (PDF) was calculated from the likelihood function assuming flat priors on the parameters, and the marginalized PDFs were converted to cumulative distribution functions to find best fit values of $z_0 = 1.57^{+0.15}_{-0.14}$ kpc and $n_0 = 0.015 \pm 0.001$ cm$^{-3}$, implying \replaced{DM$_{\perp,\infty}$}{$N_{e\perp}(\infty)$} $= 23.0 \pm 2.5$ pc cm$^{-3}$, with $\bar{\chi}^2=0.5$. The maximum log-likelihood was ${\rm ln}(\Lagr_{\rm max}) = 35.9$. Including all outliers in the fit decreased ${\rm ln}(\Lagr_{\rm max})$ to $-31.1$. The results of the PPK model are shown with observed DM$_\perp$ vs. $|z|$ in Figure~\ref{fig:scaleheightfit}, and the posterior PDF is shown in Figure~\ref{fig:triangle}. The seven outliers noted in Figure~\ref{fig:scaleheightfit} have departures from the model-predicted \replaced{DM$_\perp$}{$N_{e\perp}$} that are too large to be explained by turbulent density fluctuations, although the fluctuations could account for up to 60$\%$ of the deficient DM for J1024-0719 and up to $40\%$ of the excess DM for J0751+1807. Kolmogorov fluctuations do induce a DM variance that is consistent with the scatter in DM$_\perp$ vs. $|z|$ for most pulsars in the sample.

\indent Some previously published values of $n_0$ and $z_0$ derived from electron density  modeling of the thick disk component of the Galaxy are shown in Table~\ref{tab:prevn0}. These references were selected based on their exclusion of low latitude pulsars and pulsars with excess dispersion, similar to our own analysis. Our estimates of $n_0$, $z_0$, and \replaced{DM$_{\perp,\infty}$}{$N_{e\perp}(\infty)$} are consistent with \cite{2012MNRAS.427..664S} and \cite{2009ApJ...702.1472S}, but depart slightly from the estimates of \cite{2017ApJ...835...29Y}, \cite{2008PASA...25..184G}, and \cite{2002astro.ph..7156C}. Discrepancies in the fitting procedure of \cite{2008PASA...25..184G} have already been discussed by \cite{2009ApJ...702.1472S} and \cite{2012MNRAS.427..664S}, and likely contribute to the large contrast between their inferred scale height and more recent estimates. Our results are consistent with previous findings that the Galactic disk has a higher scale height and lower mid-plane density than those used in NE2001.

\indent The covariance between $n_0$ and $z_0$ mean that there is a range of densities and scale heights that can produce the same \replaced{DM$_{\perp,\infty}$}{$N_{e\perp}(\infty)$}.  Indeed, the most robustly determined quantity is their product, \replaced{DM$_{\perp,\infty}$}{$N_{e\perp}(\infty)$}, and the values obtained in different references are in good agreement except for \cite{2017ApJ...835...29Y}.   Our estimate of \replaced{DM$_{\perp,\infty}$}{$N_{e\perp}(\infty)$} sits about midway between the smallest and largest estimates from previous studies. The degeneracy between $n_0$ and $z_0$ can only be broken with a large distribution of accurate pulsar distance measurements across the full latitude range, with enough sampling at the highest $|z|$ to tightly constrain \replaced{DM$_{\perp,\infty}$}{$N_{e\perp}(\infty)$}. An accurate measurement of \replaced{DM$_{\perp,\infty}$}{$N_{e\perp}(\infty)$} is crucial to estimating the Galactic contribution to the DMs of extragalactic sources like FRBs. 

\begin{deluxetable}{l C R R}
\tablecaption{Selected Scale Heights and Electron Densities in the Literature \label{tab:prevn0}}
\tabletypesize{\footnotesize}
\tablewidth{0.7\textwidth}
\tablehead{\colhead{Reference} & \colhead{$n_0$ (cm$^{-3}$)} & \colhead{$z_0$ (kpc)} & \colhead{$N_{e\perp}(\infty)$}}
\startdata
    This Work & 0.015 & 1.57^{+0.15}_{-0.14} & 23.5 \pm 2.5 \\
    \cite{2017ApJ...835...29Y} & 0.011 & 1.67^{+0.05}_{-0.05} & 18.9 \pm 0.9 \\
    \cite{2012MNRAS.427..664S} & 0.015 & 1.60^{+0.30}_{-0.30} & 24.4 \pm 4.2 \\
    \cite{2009ApJ...702.1472S} & 0.016 & 1.41^{+0.26}_{-0.21} & 22.6 \pm 4.0 \\
    \cite{2008PASA...25..184G} & 0.014 & 1.83^{+0.12}_{-0.25} & 25.6 \pm 2.6 \\
    \cite{2002astro.ph..7156C} & 0.025 & \multicolumn{1}{l}{0.95} & \multicolumn{1}{l}{24.0} \\
\enddata
\tablecomments{These references were selected based on their exclusion of low latitude pulsars and pulsars with excess dispersion, similar to our own analysis. In all cases the error on $n_0$ is about $\pm 0.001$ cm$^{-3}$. The scale height and mid-plane density used in NE2001 \citep{2002astro.ph..7156C} are shown for comparison.}
\end{deluxetable}

\section{Clumps and Voids}\label{sec:outliers}

\indent Seven of the pulsar outliers identified in the fit for $n_0$ and $z_0$ and highlighted in Figure~\ref{fig:scaleheightfit} have large DM departures from the model caused by discrete clumps and voids along their LoS. Six of these notable outliers, PSRs J0922+0638 (B0919+06), J0826+2637 (B0823+26), J0751+1807, J1614$-$2230, J1623$-$0908, and J1643$-$1224, show significant excess DMs, while J1024$-$0719 shows a DM deficit. J1643$-$1224 was previously identified as lying behind the HII region Sh 2-27 \citep[e.g.,][]{2011ApJ...736...83H}. We highlight these pulsars in bold in the following sections for readers who wish to skip to discussion of specific pulsars.

\begin{figure*}[htp]
    \centering
    \includegraphics[width=0.8\textwidth]{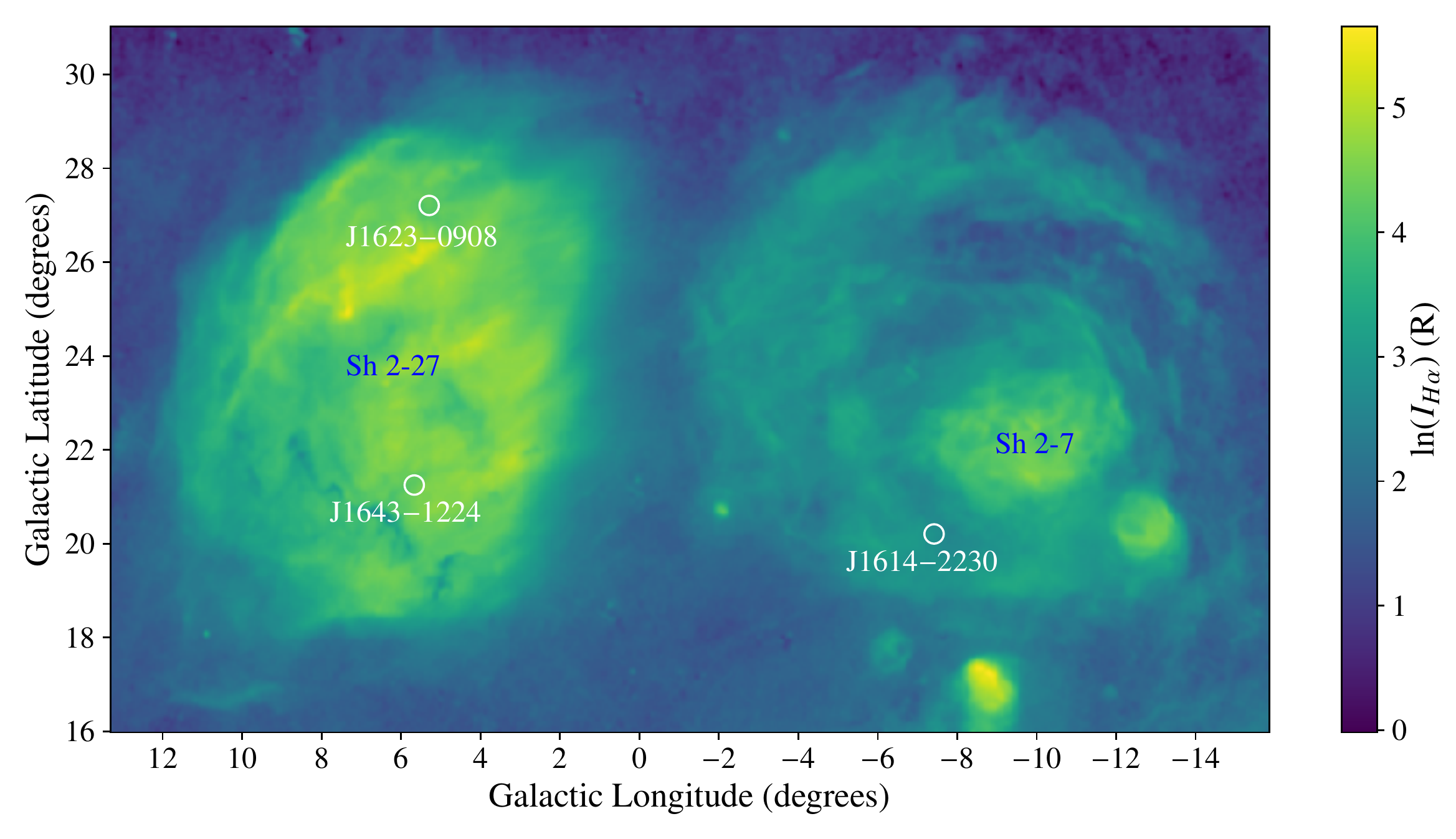}
    \caption{H$\alpha$ intensity in Rayleighs (log scale) of the HII regions in front of J1614-2230, J1623-0908, and J1643-1224 from Finkbeiner's composite of the SHASSA, VTSS, and WHAM surveys \citep{2003ApJS..146..407F}. Two regions of the map are identified in the Sharpless catalog, Sh 2-27, which was originally estimated to have an angular diameter of 8 degrees, and 2-7, which was estimated to have a diameter of 4 degrees \citep{1959ApJS....4..257S}.}
    \label{fig:h2regions}
\end{figure*}

\indent A search through H$\alpha$, radio, and molecular line surveys revealed that two other pulsars lie behind HII regions: J1623$-$0908, which also lies behind Sh 2-27, and J1614$-$2230, which lies behind a diffuse nebula containing Sh 2-7. An H$\alpha$ map of the two regions constructed using the \cite{2003ApJS..146..407F} composite of the Southern H$\alpha$ Sky Survey \citep[SHASSA;][]{2001PASP..113.1326G}, the Virginia Tech Spectral Line Survey \citep[VTSS;][]{1999AAS...195.5309D}, and the Wisconsin H$\alpha$ Mapper \citep[WHAM;][]{2003ApJS..149..405H} is shown in Figure~\ref{fig:h2regions}. We use the H$\alpha$ intensity in these pulsars' directions to place constraints on the emission measures (EMs), which we combine with scattering measures (SMs) and DMs to constrain the internal properties of the HII regions.

\indent An ionized cloudlet model \citep{1991Natur.354..121C, 1993ApJ...411..674T, 2002astro.ph..7156C} is used to relate these measures for individual HII regions.  Cloudlets have a volume filling factor $f$, internal density fluctuations with variance $\epsilon^2 = \langle (\delta n_e)^2 \rangle/{n_e}^2$, and cloud-to-cloud variations described by $\zeta = \langle {n_e}^2 \rangle/\langle {n_e} \rangle^2$. Here $n_e$ is the local, volume-averaged mean density. The EM and DM are related by 
\begin{equation}\label{eq:EMDM1}
    {\rm{EM}} = \frac{\zeta (1+\epsilon^2)}{fL} {\rm{DM}}^2,
\end{equation}
where $L$ is the depth of the region. The SM and DM are related by 
\begin{equation}\label{eq:SMDM}
\begin{split}
    {\rm{SM}} &= \left(\frac{C_{\rm SM}F_c}{L}\right) {\rm{DM}}^2\\
    &\approx (1.84 \times 10^{-6})\frac{F_c}{L_{\rm{kpc}}}{\rm{DM}}^2
\end{split}
\end{equation}
where the second line is for $F_c$ in units of pc$^{-2/3}$, $L$ in kpc, and the other quantities in standard units. The fluctuation parameter is related to the outer scale $l_0$ by $F_c = \zeta\epsilon^2/f l_0^{2/3}$, and the constant $C_{\rm SM} = 1/3(2\pi)^{1/3}$ for a Kolmogorov  spectrum. Equivalently,
\begin{equation}
    F_c \approx (5.45 \times 10^5) \times \frac{L_{\rm{kpc}}{\rm{SM}}}{{\rm{DM}}^2}
\end{equation}
for SM and DM in their usual units of kpc m$^{-20/3}$ and pc cm$^{-3}$, $L$ in kpc, and $F_c$ in pc$^{-2/3}$.\\
\indent The EMs of the HII regions are inferred from H$\alpha$ intensity ($I_{\rm{H}\alpha}$) by assuming a gas temperature $T$ and accounting for the dust extinction along the LoS,
\begin{equation}\label{eq:EMDM}
    {\rm{EM}} = 2.75 \times T_{4}^{0.9} I_{\rm{H}\alpha} e^\tau
\end{equation}
where $T_4 = T/10^4$ K, the optical depth is $\tau = 2.44E_{B-V}$ \citep[e.g.,][]{2011ApJ...736...83H}, and we assume the dust lies in front of the HII region. The color excess $E_{B-V}$ is taken from the \citet{2011ApJ...737..103S} Galactic dust extinction map.

\indent To estimate the SMs of the HII regions we assume the scattering is described by a thin screen for the HII region combined with the scattering contribution predicted by the PPK model, which we denote as SM$_{\tau-{\rm{DM}}}$ for reasons discussed below. The total SM is the sum of the SM of the HII region and the SM from the model. For a thin screen, the total SM is 
\begin{equation}
    {\rm{SM}} = {\rm{SM}}_{\rm{\tau-{\rm{DM}}}} + \int_0^d {\rm d}s \bigg(\frac{s}{d}\bigg)\bigg(1 - \frac{s}{d}\bigg) \cnsq,
\end{equation}
where $\cnsq = {\rm{SM_{HII}}}\delta(s - D_{\rm{HII}})$ \citep[e.g.,][]{2002astro.ph..7156C}. Using the model-predicted SM based on the pulsar's distance $D_\pi$, the total SM becomes
\begin{multline}
    {\rm{SM}} = {\rm{SM}}_{\rm{\tau-{\rm{DM}}}}(D_\pi)\\ + {\rm{SM_{HII}}}(D_{\rm{HII}}/D_\pi)(1 - D_{\rm{HII}}/D_\pi).
\end{multline}
The SM of the HII region is then
\begin{equation}
    {\rm{SM_{HII}}} = \frac{{\rm{SM}} - {\rm{SM}}_{\rm{\tau-{\rm{DM}}}}(D_\pi)}{(D_{\rm{HII}}/D_\pi)(1 - D_{\rm{HII}}/D_\pi)}.
\end{equation}
\indent The total SM is estimated from measurements of pulse broadening time $\tau_d$ and scintillation bandwidth $\Delta \nu_d$, and we denote this empirically estimated SM as SM$_\tau$ \citep[e.g.,][]{2002astro.ph..7156C}:
\begin{equation}\label{eq:tausm}
    \tau_d = (1.10 \hspace{0.05in} {\rm{ms}})D_\pi {\rm{SM_\tau}}^{6/5} \nu^{-22/5}
\end{equation}
\begin{equation}\label{eq:nutau}
    \Delta \nu_d = (168 \hspace{0.05in} {\rm{Hz}}) {\rm{SM_\tau}}^{-6/5} \nu^{22/5} D_\pi^{-1}.
\end{equation}
The model-predicted SM$_{\tau-{\rm{DM}}}$ is estimated using the empirically measured distribution of pulse broadening times and DMs (the $\tau_d$-DM distribution). The most recent characterization  of this distribution \citep[][]{2016arXiv160505890C} is 
\begin{multline}\label{eq:taudmdist}
    \tau_d = (2.98 \times 10^{-7} \hspace{0.05in} {\rm{ms}}) \times {\rm{DM}}^{1.4}\\\times (1 + (3.55 \times 10^{-5}\times {\rm{DM}}^{3.1}))
\end{multline}
with $\sigma_{\rm{log}\tau} = 0.76$ \citep[see also][]{2015ApJ...804...23K}. The $\tau_d$-DM distribution is evaluated at the PPK model's prediction of the pulsar DMs to yield a prediction of $\tau_d$ that is converted into SM$_{\tau-{\rm{DM}}}$ using Eq.~\ref{eq:tausm}.

\subsection{The HII Region Sh 2-27}\label{sec:Sh227}
\indent The HII region \object{Sh 2-27} is energized by the O-star $\zeta$ Oph at a distance of $112^{+3}_{-2}$ pc \citep{2007A&A...474..653V}. PSRs \textbf{J1623$-$0908} and \textbf{J1643$-$1224} are in the directions ($5.30^\circ$, $27.18^\circ$) and ($5.67^\circ$, $21.22^\circ$), respectively, and are seen through this region. Since the Finkbeiner H$\alpha$ map gives the total H$\alpha$ intensity on the sky, to obtain the H$\alpha$ intensity of the HII region we subtract a mean background intensity found by averaging the intensities of pixels directly around the HII region. This background subtraction yields an H$\alpha$ intensity of $67.9 \pm 4.9$ Rayleighs for J1623$-$0908 and $69.5 \pm 4.6$ R for J1643$-$1224, with quoted errors from the associated intensity error map. The color excess for J1623$-$1908 is $E_{B-V} = 0.272 \pm 0.009$ and for J1643$-$1224 it is $E_{B-V} = 0.377 \pm 0.022$ \citep{2011ApJ...737..103S}. Assuming a gas temperature of 7000 K and correcting for extinction yields two values of the EM from the HII region: $263 \pm 25$ pc cm$^{-6}$ in the direction of J1623$-$0908 and $348 \pm 44$ pc cm$^{-6}$ in the direction of J1643$-$1224. Using EM we estimate the mean electron density inside the HII region, assuming the cloudlets are homogeneously distributed and the region is as wide as it is deep: 
\begin{equation}\label{eq:neapprox}
    n_{e,{\rm{HII}}} \approx \left[ \frac{\rm{EM}}{\zeta (1+\epsilon^2) fL}\right]^{1/2}
\end{equation}
where $f$ is the filling factor and $L$ is the depth of the region.  For given EM and $L$
the maximum possible density is for $\zeta = 1+\epsilon^2 = 1$,
\begin{equation}
n_{e,{\rm{HII}}}^{\rm(max)} \approx \left( \frac{\rm{EM}}{ fL}\right)^{1/2}.
\end{equation}
\citet[][]{2011ApJ...736...83H} measure the diameter of Sh 2-27 to be $L = 34$ pc and assume $f = 0.1$, which yields $n_{e,{\rm{HII}}}^{\rm(max)} = 8.8 \pm 1.3$ cm$^{-3}$ in the direction of J1623$-$0908 and $n_{e,{\rm{HII}}}^{\rm(max)} = 10.1 \pm 1.1$ cm$^{-3}$ in the direction of J1643$-$1224. Taking DM$_{{\rm{HII}}} \approx f n_{e,{\rm{HII}}} L$ then gives an estimated DM enhancement for J1623$-$0908 of $29.9 \pm 4.3$ pc cm$^{-3}$ and $34.4 \pm 4.5$ pc cm$^{-3}$ for J1643$-$1224, assuming $\zeta = (1+\epsilon^2) = 1$. The density and DM can be significantly smaller for other values of $\zeta$ and $\epsilon$.

\indent However, the DM excesses implied by the PPK model are much higher: $48.6 \pm 1.5$ pc cm$^{-3}$ for J1623$-$0908 and $56.2 \pm 0.4$ pc cm$^{-3}$ for J1643$-$1224. If all of the excess DM is from the HII region, then either the filling factor, the temperature, and/or the depth of the region must be larger than \cite{2011ApJ...736...83H} assume. Assuming that the excess DM is from the HII region, combining EM and DM using Eq.~\ref{eq:EMDM} for both J1623$-$0908 and J1643$-$1224 yield the same result assuming $L = 34$ pc: $f/\zeta (1 + \epsilon^2) = 0.27 \pm 0.02$. This estimate does not account for any error in $L$. \cite{2011ApJ...736...83H} estimate $L$ by carefully mapping the boundary of the region in H$\alpha$ and assuming the HII region is a sphere, but they do not provide the uncertainty in this estimate. Therefore the error quoted on $f/\zeta (1 + \epsilon^2)$ should only be taken as a rough estimate.

\indent There is disagreement between timing parallaxes measured for J1643$-$1224; \cite{2016MNRAS.458.3341D} measure $1.17 \pm 0.26$ mas, \cite{2016MNRAS.455.1751R} measure $1.27 \pm 0.19$ mas, and \cite{2009MNRAS.400..951V} measure $2.2 \pm 0.4$ mas. We use the latter due to the more conservative treatment of timing red noise, but using the other values of parallax would modestly change the DM excess from the model to 51--52 pc cm$^{-3}$, making $f/\zeta(1+\epsilon^2) \approx 0.22-0.23$. An interferometric parallax for this pulsar, unaffected by red timing noise, will be forthcoming from the next release of PSR$\pi$ \citep[][and personal communication]{2019ApJ...875..100D}, and may alter our estimate of the DM contribution of Sh 2-27.

\indent We estimate the SM of Sh 2-27 using the scintillation bandwidth for J1643$-$1224, $\Delta \nu_d = 0.022$ MHz at 1.5 GHz \citep{2013MNRAS.429.2161K}, which   gives SM$_\tau$ = 0.15 kpc m$^{-20/3}$ from Eq.~\ref{eq:nutau}. The PPK model predicts a DM of $6.3 \pm 0.5$ pc cm$^{-3}$ for this pulsar's distance and latitude. Evaluating the empirical $\tau_d$-DM distribution (c.f. Eq.~\ref{eq:taudmdist}) at this DM and converting to SM$_{\tau-{\rm{DM}}}$ using Eq.~\ref{eq:tausm} yields SM$_{\tau-{\rm{DM}}} = (5.7 \pm 3.9) \times 10^{-5}$ kpc m$^{-20/3}$. For Sh 2-27 we thus find SM$_{\rm{HII}} = 0.8$ kpc m$^{-20/3}$. The fluctuation parameter is $F_c = 4.6$ pc$^{-2/3}$. These values should be taken as rough estimates since the errors depend in part on the uncertainty of $\Delta \nu_d$ and $L$, both of which are unstated by their respective sources. Nonetheless, $F_c$ is much larger than expected far above the Galactic plane, and is similar to values estimated for regions of high stellar activity in the spiral arms \citep{1991Natur.354..121C, 2002astro.ph..7156C}. 

\subsection{The HII Region Sh 2-7}\label{sec:Sh27}
The HII region \object{Sh 2-7} lies within a diffuse nebula at a distance of 0.17 kpc \citep{1974A&AS...16..163S, 1980A&A....92..156B}. Based on its angular size and distance, Sh 2-7 is estimated to be 21 by 12 parsecs face-on \citep{1980A&A....92..156B}. PSR \textbf{J1614$-$2230} is in the direction $(352.64^\circ, 20.19^\circ)$ at a distance of $0.77^{+0.06}_{-0.05}$ kpc \citep{2016AA...587A.109G} and is seen through this region. The background-subtracted H$\alpha$ intensity in the pulsar's direction is $I_{\rm{H}\alpha} = 15.5 \pm 2.1$ R, and the color excess is $E_{B-V} = 0.213 \pm 0.003$. Assuming a gas temperature of 7000 K yields an EM of $52.0 \pm 7.5$ pc cm$^{-6}$. Approximating the shape of the region to be $L \approx \sqrt{21\times 12} = 16$ pc,
and $f = 0.1$ yields $n_{e,{\rm{HII}}}^{\rm(max)} = 5.7 \pm 0.4$ cm$^{-3}$, and DM$_{{\rm{HII}}} \approx fn_{e,{\rm{HII}}}L = 9.1 \pm 0.6$ pc cm$^{-3}$. The DM excess predicted by the PPK model is $24.4 \pm 0.7$ pc cm$^{-3}$, indicating again that $f$, $L$, and/or $T$ are higher in this region. Assuming this excess is due to the HII region, Eq.~\ref{eq:EMDM} implies that $f/\zeta (1 + \epsilon^2) = 0.71 \pm 0.09$, assuming $L = 16$ pc.

\indent We estimate the SM of Sh 2-7 using the scintillation bandwidth of J1614$-$2230, which is $\Delta \nu_d = 4.9 \pm 0.5$ MHz at a reference frequency of 1.5 GHz \citep{2020ApJ...890..123S}, implying SM$_\tau$ = $(1.04 \pm 0.14) \times 10^{-3}$ kpc m$^{-20/3}$. Performing the same analysis as for Sh 2-27 yields SM$_{\rm{HII}} = 0.0058$ kpc m$^{-20/3}$ and $F_c = 0.08$ pc$^{-2/3}$. These values should again be taken as estimates since the uncertainty in the distance and size of the HII region is unknown. The estimated value of $F_c$ is much closer to the expected values for low density regions in the local ISM \citep{2002astro.ph..7156C}, which may be related to the lower intensity, more diffuse H$\alpha$ emission seen in this region.

\subsection{The Superbubble GSH 238+00+09}
\indent PSR \textbf{J1024$-$0719} is in the direction $(251.702^\circ, 40.515^\circ)$ at a distance of 1.12 kpc \citep{2016AA...587A.109G}, and exhibits a DM deficit from the PPK model of $6.12 \pm 0.86$ pc cm$^{-3}$. \cite{1999ApJ...523L.171T} use the Schlovskii effect to infer an upper limit distance to J1024$-$0719 of 226 pc, resulting in a lower limit to the mean electron density along this LoS of $n_e > 0.029$ cm$^{-3}$. They interpret this mean density as an excess resulting from an interaction between the Local Hot Bubble and the superbubble \object{GSH 238+00+09}, discovered by \cite{1998ApJ...498..689H}. The more recent distance derived from parallax is more than four times larger than the value quoted by \citep{1999ApJ...523L.171T}, and combined with DM $= 6.477$ pc cm$^{-3}$ implies a mean electron density $n_e \approx 0.006$ cm$^{-3}$, much lower than the density predicted by the PPK model. Furthermore, the revised distance places J1024$-$0719 not at the edge of the superbubble, but well within it or even beyond it.

\indent This superbubble was first identified by \cite{1998ApJ...498..689H} in HI, IR, $\sfrac{1}{4}$-keV X-ray, and radio observations as an approximately 0.5 kpc long cavity centered on a longitude of $238^\circ$ at a central distance of about 0.8 kpc, with an angular radius in the Galactic plane of about $16^\circ \approx 224$ pc. The superbubble has since been observed in differential color excess maps derived from stellar extinction \citep{2014A&A...566A..13P, 2015MmSAI..86..626L}. \cite{2014A&A...566A..13P} combine a 3D dust map with soft X-ray data to find that the superbubble is likely filled with hot ($\sim 10^6$ K) gas, although CaII absorption observed through the cavity suggests that it may be partially filled with warm ionized gas with a filling factor less than 1. This more recent dust extinction map suggests that the superbubble may be larger than \cite{1998ApJ...498..689H} originally inferred, extending from about 0.2 kpc to 1.2 kpc away from the Sun, and as wide as about 300 pc. This revised size suggests that J1024-0719 lies well within the cavity or beyond it, depending on how the shape of the cavity changes at higher latitudes.

\indent We estimate the mean electron density inside the bubble assuming the medium extending to the superbubble is still consistent with the PPK model. The PPK model predicts that up to the edge of the superbubble at $D = 0.2$ kpc and a latitude of $40.5^\circ$, the DM will be $2.9 \pm 0.5$ pc cm$^{-3}$. If the rest of the pulsar's DM is contributed by electrons in the bubble, then the bubble contributes a DM of $3.6 \pm 0.5$ pc cm$^{-3}$ over a distance of $0.92$ kpc. This gives a lower limit on the mean electron density in the bubble of $n_e \geq 0.004 \pm 0.001$ cm$^{-3}$, assuming the filling factor $f \leq$ 1, and that the cavity is as long at $b = 40.5^\circ$ as it is in the Galactic plane. Such a low density is consistent with the cavity being filled by hot gas ($T \sim 10^6$ K).

\indent Assuming the superbubble is in pressure equilibrium with the warm ionized medium (WIM) surrounding it, the electron number density inside the bubble is roughly $n_{e,{\rm bubble}} \approx n_{e,{\rm WIM}}(T_{e,{\rm WIM}}/T_{e,{\rm bubble}})$. Adopting typical values of $10^4$ K for the WIM and $10^6$ K for the hot, X-ray emitting gas, $n_{e,{\rm bubble}}$ should be about two orders of magnitude smaller than $n_{e,{\rm WIM}}$. The PPK model implies WIM electron densities $\sim 10^{-2}$ cm$^{-3}$ and a mean density in the bubble $>0.004$ cm$^{-3}$, slightly larger than the density implied by pressure equilibrium. This estimate is overly simplistic, since \cite{1998ApJ...498..689H} finds evidence of expansion from the velocity of LSR inferred from HI observations. Moreover, \cite{2014A&A...566A..13P} find evidence for some warm ionized gas within the superbubble, in addition to the hot gas responsible for the observed soft X-ray emission. \cite{1998ApJ...498..689H} uses $\sfrac{1}{4}$ keV and $\sfrac{3}{4}$ keV ROSAT X-ray observations to infer $T_{e,{\rm bubble}} \approx 1$-2 $\times 10^6$ K, and converting to EM estimates the mean electron density to be $n_{e,{\rm bubble}} \approx 0.007$ cm$^{-3}$, which is consistent with our result.

\indent The other pulsars J0751+1807, J0922+0638 (B0919+06), and J0826+2637 (B0823+26) have anomalously high DMs that might also be related to this superbubble. \cite{2001ApJ...550..287C} use interstellar scintillometry of B0919+06 and B0823+26 to infer the presence of a turbulent interface region between the Local Bubble and GSH 238+00+09, which is consistent with previous predictions of clumps of dense, ionized material left between the two structures during their formation. Indeed, \cite{2014A&A...566A..13P} find a dense cloud implied by excess dust opacity immediately between the Local Bubble and the superbubble, which attenuates some of the X-ray emission in that direction. As \cite{1999ApJ...523L.171T} suggest, the excess electron density observed towards B0919+06 and B0823+26 could indicate a large density gradient near the edge of the superbubble. J0751+1807 is much farther than these pulsars (1.5 kpc), but some of its excess DM could be contributed by the turbulent, dense interface. If this is the case, then the superbubble (and even the Local Bubble) extend to large Galactic latitudes (as \cite{2001ApJ...550..287C} note).

\begin{deluxetable*}{c R R C R C C C C}
\tabletypesize{\footnotesize}
\tablewidth{\textwidth}
\tablecaption{Constraints on Clump Properties \label{tab:clumps}}
\tablehead{\colhead{Pulsar} & \colhead{$l$} & \colhead{$b$} & \colhead{DM} & \colhead{$D$} & \colhead{DM $-$ DM$_{\rm{PPK}}$} & \colhead{$I_{\rm{H}\alpha}$} & \colhead{EM} & \colhead{$L_{\rm m} [gf/\zeta(1+\epsilon^2)]$}\\
\colhead{} & \multicolumn{2}{c}{(Degrees)} & \colhead{(pc cm$^{-3}$)} & \colhead{(kpc)} & \colhead{(pc cm$^{-3}$)} & \colhead{(R)} & \colhead{(pc cm$^{-6}$)} & \colhead{(pc)} }
\tablecolumns{9}
\startdata
     J0437$-$4715$^\dagger$ & 253.39 & -41.96 & 2.65 & 0.156^{+0.001}_{-0.001} & 0.42 \pm 0.13 & <1.7 & <3.5 & >0.05 \\
     J0751+1807$^\dagger$ & 202.73 & 21.09 & 30.25 & 1.51^{+0.45}_{-0.28} & 11.3 \pm 1.3 & 3.3 \pm 0.6 & 7.3 \pm 1.3 & 17 \pm 4  \\
     B0823+26$^\dagger$ & 196.96 & 31.74 & 19.48 & 0.497^{+0.002}_{-0.003} & 12.7 \pm 0.4 & 2.0 \pm 0.5 & 4.5 \pm 1.0 & 36 \pm 7 \\ 
     B0919+06$^\dagger$ & 225.42 & 36.39 & 27.30 & 1.20^{+0.22}_{-0.16} & 13 \pm 1 & 2.4 \pm 0.5 & 5.5 \pm 1.2 & 30 \pm 6 \\
     J1614$-$2230$^\dagger$ & 352.64 & 20.19 & 34.92 & 0.77^{+0.06}_{-0.05} & 24.4 \pm 0.7 & 15.5 \pm 2.1 & 52.0 \pm 7.1 & 11 \pm 1\\
     J1623$-$0908$^\dagger$ & 5.30 & 27.18 & 68.18 & 1.69^{+0.34}_{-0.24} & 48.6 \pm 1.5 & 67.9 \pm 4.9 & 263 \pm 19 & 9.0 \pm 0.1 \\
     J1643$-$1224$^\dagger$ & 5.67 & 21.22 & 62.41 & 0.45^{+0.10}_{-0.07} & 56.2 \pm 0.4 & 69.5 \pm 4.6 & 348 \pm 23 & 9.1 \pm 0.6 \\
     B1839+56 & 86.08 & 23.82 & 26.77 & 1.52^{+0.02}_{-0.13} & 8.4 \pm 1.3 & 2.3 \pm 0.5 & 5.2 \pm 1.1 & 14 \pm 4 \\
     B2003$-$08 & 34.10 & -20.30 & 32.39 & 2.35^{+0.74}_{-0.05} & 5.3 \pm 2.1 & 2.8 \pm 0.5 & 7.3 \pm 1.3 & 4 \pm 2 \\
     J2129$-$5721 & 338.01 & -43.56 & 31.85 & 2.36^{+0.62}_{-0.40} & 10 \pm 2 & < 2.2 & < 4.8 & > 22\\
     J2144$-$3933 & 2.79 & -49.47 & 3.35 & 0.16^{+0.02}_{-0.01} & 1.0 \pm 0.1 & < 1.4 & < 2.9 & > 0.3 \\
     B2303+30 & 97.72 & -26.66 & 49.58 & 4.48^{+0.64}_{-0.58} & 12 \pm 4 & 2.3 \pm 0.5 & 5.4 \pm 1.1 & 29 \pm 13 \\
\enddata
\tablecomments{Pulsars that have complementary information about their LoS are indicated with a $\dagger$. J0751+1807, B0823+26, and B0919+06 are discussed in Section~\ref{OICsection}; J1614$-$2230, J1623$-$0908, and J1643$-$1224 are discussed in Sections~\ref{sec:Sh227} and~\ref{sec:Sh27}; and J0437$-$4715 is discussed in Appendix~\ref{app:outliers}. The remaining pulsars do not have additional information about their LoS; brief discussion of these pulsars is in Appendix~\ref{app:outliers}. DM $-$ DM$_{\rm{PPK}}$ refers to the observed DM minus the DM predicted by the PPK model, which we interpet as the clump DM. The errors quoted on EM (and subsequently $L_{\rm m} [gf/\zeta(1 + \epsilon^2)]$) are based on the \cite{2003ApJS..146..407F} $I_{\rm{H}\alpha}$ error map, but these errors should be interpreted as estimates since they are based on a composite of the errors in the WHAM, SHASSA, and VTSS maps and do not account for all of the covariances. For J0437$-$4715, J2129$-$5721, and J2144$-$3933, $I_{\rm{H}\alpha}$ is an upper limit. $L_{\rm m} [gf/\zeta(1 + \epsilon^2)]$ denotes the characteristic size of the clump weighted by the fraction $g$ of EM that the clump contributes, the filling factor $f$, the cloud-to-cloud variations $\zeta$, and internal density fluctuation variance $\epsilon^2$. If $L_{\rm m} [gf/\zeta(1 + \epsilon^2)] > 1$, then the maximum clump size possible, $L_{\rm m} = L_{\rm max}$, corresponds to $g = f = \zeta = 1$ and $\epsilon = 0$. If $L_{\rm m} [gf/\zeta(1 + \epsilon^2)] < 1$, then $L_{\rm m} = L_{\rm min}$ for $g = f = \zeta = 1$ and $\epsilon = 0$.}
\end{deluxetable*}

\subsection{Optically Invisible Clumps}\label{OICsection}
Nine of the pulsars in our sample have DMs in excess of the PPK model prediction, but do not show enough H$\alpha$ emission to indicate visibly dense HII regions along the LoS. Nonetheless, we use the H$\alpha$ intensity at these locations on the sky and excess DM implied by the PPK model to place constraints on the ionized clumps of gas causing this excess DM. The EMs and excess DMs for these pulsars are listed in Table~\ref{tab:clumps}. Assuming a fraction $g$ of the EM and all of the excess DM arise from a single clump of ionized gas along the LoS, the characteristic size of this clump, weighted by the filling factor, cloud-to-cloud variance, and internal density fluctuations, is given by DM$^2$/EM = $L_{\rm m} [gf/\zeta(1 + \epsilon^2)]$ (c.f. Eq.~\ref{eq:EMDM1}). The maximum depth of the clump $L_{\rm m} = L_{\rm max}$ corresponds to $g = f = \zeta = 1$ and $\epsilon = 0$ if $L_{\rm m} [gf/\zeta(1 + \epsilon^2)] > 1$. If $L_{\rm m} [gf/\zeta(1 + \epsilon^2)] < 1$, then $L_{\rm m} = L_{\rm min}$, the minimum depth of the clump. The constraints on the characteristic depth $L_{\rm m} [gf/\zeta(1 + \epsilon^2)]$ based on H$\alpha$ intensity and excess DM for each LoS are shown in Table~\ref{tab:clumps}. In this section, we summarize the implications for three of these pulsars, J0751+1807, B0919+06, and B0823+26, given the proximity of their LoS to the superbubble GSH 238+00+09. The rest of the pulsars shown in Table~\ref{tab:clumps} are discussed in Appendix~\ref{app:outliers}.

\indent \textbf{J0751+1807} has an excess DM of $11.3 \pm 1.3$ pc cm$^{-3}$ and clump parameter $L_{\max} [gf/\zeta(1 + \epsilon^2)] = 17 \pm 4$ pc. Since J0751+1807 has a longitude of $l = 202.73^\circ$ and distance $D = 1.5$ kpc, it is possible that some or all of this excess DM is the result of viewing the pulsar through a dense, turbulent interface at the edge of the superbubble GSH 238+00+09, similar to B0823+26 and B0919+06.  

\indent \textbf{J0826+2637 (B0823+26)} has a DM excess equal to $12.7 \pm 0.4$ pc cm$^{-3}$ and clump parameter $L_{\max} [gf/\zeta(1 + \epsilon^2)] = 36 \pm 7$ pc. The pulsar's location and DM excess are also consistent with a dense region near the edge of the superbubble, as was first noted by \cite{1999ApJ...523L.171T}.  

\indent \textbf{J0922+0638 (B0919+06)} has a DM excess of $13 \pm 1$ pc cm$^{-3}$. From the excess DM and H$\alpha$ emission we estimate $L_{\max} [gf/\zeta(1 + \epsilon^2)] = 30 \pm 6$ pc, similar to B0823+26. The pulsar's longitude $l = 225.42^\circ$ and distance $D = 1.2$ kpc suggest that it is also viewed through a dense region near the edge of the superbubble, similar to J0751+1807 and B0823+26. Using scintillometry, \cite{2001ApJ...550..287C} estimate a scattering screen $<300$ pc away, but our limit on $L_{\max} [gf/\zeta(1 + \epsilon^2)]$ implies $L$ is on the order of 10s of parsecs, larger than most screen sizes inferred from scintillation.

\indent The values of the composite parameter for B0823+26 and B0919+06 are strikingly similar, despite their large contrast in distance. However, the interplay between the five parameters $g$, $f$, $L$, $\zeta$, and $\epsilon$ mean that the clumps causing excess DM in these pulsars could take on a variety of forms. The factors $g$ and $\epsilon$ have a range of possible values from 0 to 1 and $\zeta \geq 1$, while $f$ is likely between 0.1 and 0.4 (for HII gas). Accordingly, the clumps could be compact in size and could contribute most of the H$\alpha$ emission observed, taking up $\sim 10$ parsecs but with high filling factors $f$ and high values of $g$. Conversely, they could be very diffuse, with large depths $L$ and low $f$. More stringent constraints on the clumps' properties require complementary observations, such as scintillation arcs, which can constrain $L$ \citep[e.g.,][]{2005ApJ...619L.171H, 2010ApJ...708..232B}. 

\section{Extrapolation to Lower Galactic Latitudes}

\begin{figure*}[htp]
    \centering
    \includegraphics[width=0.7\textwidth]{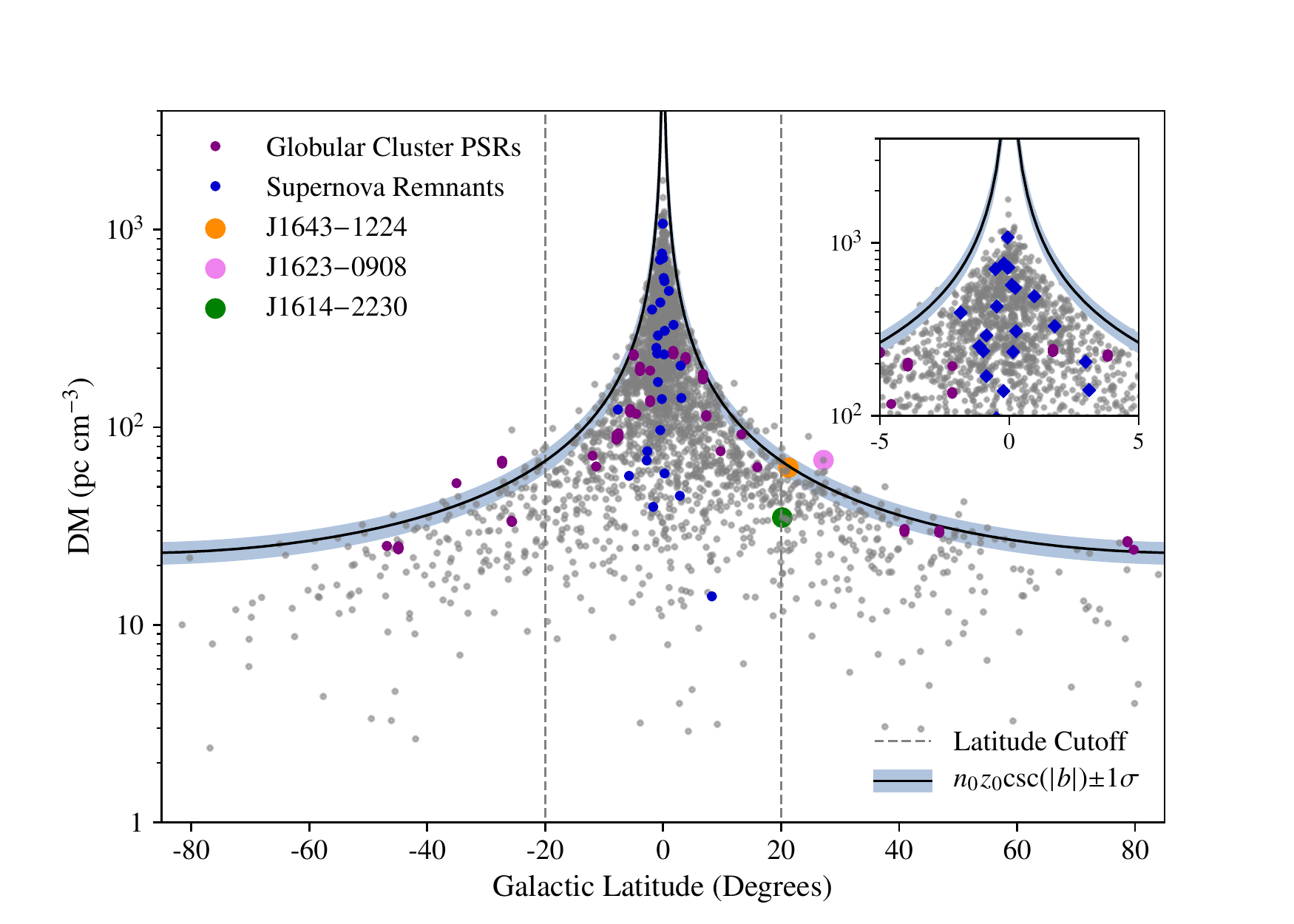}
    \caption{DM vs. Galactic latitude for all Galactic pulsars in the ATNF Pulsar Catalogue \citep{2005AJ....129.1993M}. Pulsars associated with supernova remnants are highlighted in blue and pulsars associated with globular clusters in purple. The three pulsars in our sample located behind HII regions, J1643$-$1224, J1623$-$0908, and J1614$-$2230, are also noted. The maximum Galactic DM contribution ($n_0z_0$csc($|b|$)) given by the plane-parallel component of the electron density model is shown in black, along with the corresponding $\pm 1\sigma$ error. The model was fit using the latitude cutoff shown, $|b|>20^\circ$. An inset shows pulsar DMs for latitudes $-5^\circ < b < 5^\circ$. }
    \label{fig:dmvsb}
\end{figure*}

\added{The electron density model described in this paper only applies to the thick disk of the Galaxy. Models of the entire Galaxy include both thin and thick disk components, with the thin disk consisting of a smaller scale height and larger electron density extending just a few degrees in latitude above the Galactic plane \citep[e.g.,][]{2002astro.ph..7156C, 2017ApJ...835...29Y}. To avoid the large number of supernova remnants, HII regions, and other complex features close to the Galactic plane, we limited our pulsar sample to those above $|20|^\circ$ Galactic latitude, but the thick disk likely extends down to lower latitudes. The distribution of all Galactic pulsar DMs and latitudes listed in the Australia Telescope National Facility (ATNF) Pulsar Catalogue \citep{2005AJ....129.1993M}\footnote{\url{http://www.atnf.csiro.au/research/pulsar/psrcat}} is shown in Figure~\ref{fig:dmvsb}. The PPK model estimate of the maximum Galactic DM contribution for a given latitude, $n_0 z_0$csc($|b|$), is largely consistent with the observed DM vs. $b$ distribution down to $|b| \approx 5^\circ$, with a few exceptions: A number of pulsars have DMs larger than $n_0 z_0$csc($|b|$), which likely are due to additional clumps along the LoS (e.g., J1643$-$1224 lies behind an HII region and has a DM $>n_0 z_0$csc($|b|$)). Many pulsars with DMs $<n_0 z_0$csc($|b|$) have supernova remnant associations or HII regions along their LoS (e.g., see J1614$-$2230 in Figure~\ref{fig:dmvsb}), also indicating departures from a plane-parallel medium.}

\indent \added{ Due to the unrealistic behavior of the plane-parallel model near $b=0^\circ$, more realistic Galactic models like NE2001 adopt a thick disk that rolls off with Galactocentric radius. Below latitudes of $|b|\approx5^\circ$, the plane-parallel medium predicts a maximum Galactic DM contribution consistently larger than pulsar DMs in that latitude range, suggesting the thin disk dominates LoS near the Galactic plane. Assuming the thick disk has a radius of about 15 kpc and a constant mid-plane density $n_0 \approx 0.015$ cm$^{-3}$, a distant object on the opposite side of the Galactic center would have a DM $\sim 350$ pc cm$^{-3}$ (ignoring any radial dependence or clumps and voids). Not only is this DM much smaller than DMs observed near $b = 0^\circ$, but the highest DMs observed are from pulsars near the Galactic Center, not near the opposite edge of the disk. At these low latitudes ($|b| \lesssim 5^\circ$), an additional thin disk component that rolls off with a shallower scale height and higher mid-plane density is needed. Inclusion of the thick disk model described here in full Galactic electron density models will therefore require careful consideration of the appropriate latitude ranges in which the thin and thick disk components dominate the electron density contribution to DM. While the thick disk model is likely applicable down to latitudes of about $\pm 5^\circ$, the number of clumps and voids in the local ISM increases greatly at these lower latitudes and will also require consideration.}

\section{Summary and Conclusions}
\indent We construct a model for the \replaced{Galactic electron density at high Galactic latitudes}{electron density of the thick disk of the Galaxy} that adds density fluctuations following a Kolmogorov wavenumber spectrum to the plane-parallel medium. Using the most updated sample of publicly available pulsar distance measurements \added{at Galactic latitudes greater than $20^\circ$}, obtained both from parallaxes and globular cluster associations, we find a scale height $z_0=1.57^{+0.15}_{-0.14}$ kpc and a mid-plane density $n_0=0.015 \pm 0.001$ cm$^{-3}$. The scatter in the \added{observed} distribution of DM$_\perp$($z$) is consistent with turbulent density fluctuations following a wavenumber spectrum. Nonetheless, there are several pulsars at these high latitudes that show significant departures from the model, suggesting the presence of discrete inhomogeneities (clumps and voids). Two of these pulsars, J1623$-$0908 and J1643$-$1224, lie behind the HII region Sh 2-27, and the pulsar J1614$-$2230 lies behind a diffuse nebula containing Sh 2-7. We estimate the internal density and filling factor of these HII regions by combining the measured H$\alpha$ intensity of these regions with the pulsars' model-predicted DMs. For both regions we find that the filling factors are likely $>0.1$. Four other pulsars, J1024$-$0719, J0751+1807, B0919+06, and B0823+26, have DM departures that may be attributed to the superbubble GSH 238+0+09 in the third Galactic quadrant. J1024$-$0719 lies at a distance and longitude consistent with it inhabiting the superbubble, and it exhibits a DM deficit from the model that is consistent with the bubble full of hot ionized gas, as suggested by observed X-ray emission \citep{2014A&A...566A..13P}. J0751+1807, B0919+06, and B0823+26 exhibit significant excess DM from the model, and lie at distances and longitudes that suggest they are seen through a dense interface at the edge of the superbubble, as was originally suggested by \cite{1999ApJ...523L.171T} and \cite{2001ApJ...550..287C}.\\
\indent These findings paint a picture of the ISM in the Galactic disk that significantly complicates previous assumptions of a plane-parallel medium. In addition to turbulent density fluctuations, there are discrete clumps and voids ranging widely in size and internal density. Accurate predictions of pulsar and FRB distances based on DM and electron density models must account for these complexities, which will likely grow in number as more pulsar distance measurements are obtained at high latitudes. Pulsar surveys will dramatically increase the number of known pulsars to $\sim 10^4$ in the coming decade or two, which will allow us to both more accurately map known HII regions and voids and better constrain structures causing the enhanced DMs of pulsars farthest from the Galactic plane, including those located in globular clusters. Updating constraints on the Galactic disk contribution to FRB DMs will likewise improve our understanding of the DM contributions of the Galactic halo, intergalactic medium, and FRB host galaxies. The revised electron density scale height and mid-plane density reported here \added{for the thick disk}, along with the discrete inhomogeneities described, should be incorporated into the next generation of full Galactic electron density models. 

\acknowledgements
We thank the anonymous reviewer, Michael Lam, and Joseph Lazio for their helpful suggestions. The authors acknowledge support from the National Science Foundation (NSF AAG-1815242) and are members of the NANOGrav Physics Frontiers Center, which is supported by the NSF award PHY-1430284. The Virginia Tech Spectral-Line Survey (VTSS), the Southern H-Alpha Sky Survey Atlas (SHASSA), and the Wisconsin H-Alpha Mapper (WHAM) are all funded by the NSF. SHASSA observations were obtained at Cerro Tololo Inter-American Observatory, which is operated by the Association of Universities for Research in Astronomy, Inc., under cooperative agreement with the NSF.

\bibliography{bib}

\begin{thebibliography}{}
\expandafter\ifx\csname natexlab\endcsname\relax\def\natexlab#1{#1}\fi
\providecommand{\url}[1]{\href{#1}{#1}}
\providecommand{\dodoi}[1]{doi:~\href{http://doi.org/#1}{\nolinkurl{#1}}}
\providecommand{\doeprint}[1]{\href{http://ascl.net/#1}{\nolinkurl{http://ascl.net/#1}}}
\providecommand{\doarXiv}[1]{\href{https://arxiv.org/abs/#1}{\nolinkurl{https://arxiv.org/abs/#1}}}

\bibitem[{{Armstrong} {et~al.}(1995){Armstrong}, {Rickett}, \&
  {Spangler}}]{1995ApJ...443..209A}
{Armstrong}, J.~W., {Rickett}, B.~J., \& {Spangler}, S.~R. 1995, ApJ, 443, 209,
  \dodoi{10.1086/175515}

\bibitem[{{Baart} {et~al.}(1980){Baart}, {de Jager}, \&
  {Mountfort}}]{1980A&A....92..156B}
{Baart}, E.~E., {de Jager}, G., \& {Mountfort}, P.~I. 1980, AAP, 92, 156

\bibitem[{{Bell} {et~al.}(1995){Bell}, {Bailes}, {Manchester}, {Weisberg}, \&
  {Lyne}}]{1995ApJ...440L..81B}
{Bell}, J.~F., {Bailes}, M., {Manchester}, R.~N., {Weisberg}, J.~M., \& {Lyne},
  A.~G. 1995, ApJL, 440, L81, \dodoi{10.1086/187766}

\bibitem[{{Bhat} {et~al.}(1998){Bhat}, {Gupta}, \& {Rao}}]{1998ApJ...500..262B}
{Bhat}, N.~D.~R., {Gupta}, Y., \& {Rao}, A.~P. 1998, ApJ, 500, 262,
  \dodoi{10.1086/305715}

\bibitem[{{Bhat} {et~al.}(2016){Bhat}, {Ord}, {Tremblay}, {McSweeney}, \&
  {Tingay}}]{2016ApJ...818...86B}
{Bhat}, N.~D.~R., {Ord}, S.~M., {Tremblay}, S.~E., {McSweeney}, S.~J., \&
  {Tingay}, S.~J. 2016, ApJ, 818, 86, \dodoi{10.3847/0004-637X/818/1/86}

\bibitem[{{Brisken} {et~al.}(2002){Brisken}, {Benson}, {Goss}, \&
  {Thorsett}}]{2002ApJ...571..906B}
{Brisken}, W.~F., {Benson}, J.~M., {Goss}, W.~M., \& {Thorsett}, S.~E. 2002,
  ApJ, 571, 906, \dodoi{10.1086/340098}

\bibitem[{{Brisken} {et~al.}(2010){Brisken}, {Macquart}, {Gao}, {Rickett},
  {Coles}, {Deller}, {Tingay}, \& {West}}]{2010ApJ...708..232B}
{Brisken}, W.~F., {Macquart}, J.~P., {Gao}, J.~J., {et~al.} 2010, ApJ, 708,
  232, \dodoi{10.1088/0004-637X/708/1/232}

\bibitem[{{Brownsberger} \& {Romani}(2014)}]{2014ApJ...784..154B}
{Brownsberger}, S., \& {Romani}, R.~W. 2014, ApJ, 784, 154,
  \dodoi{10.1088/0004-637X/784/2/154}

\bibitem[{{Chatterjee} {et~al.}(2001){Chatterjee}, {Cordes}, {Lazio}, {Goss},
  {Fomalont}, \& {Benson}}]{2001ApJ...550..287C}
{Chatterjee}, S., {Cordes}, J.~M., {Lazio}, T.~J.~W., {et~al.} 2001, ApJ, 550,
  287, \dodoi{10.1086/319735}

\bibitem[{{Chatterjee} {et~al.}(2009){Chatterjee}, {Brisken}, {Vlemmings},
  {Goss}, {Lazio}, {Cordes}, {Thorsett}, {Fomalont}, {Lyne}, \&
  {Kramer}}]{2009ApJ...698..250C}
{Chatterjee}, S., {Brisken}, W.~F., {Vlemmings}, W.~H.~T., {et~al.} 2009, ApJ,
  698, 250, \dodoi{10.1088/0004-637X/698/1/250}

\bibitem[{{Cordes} \& {Lazio}(2002)}]{2002astro.ph..7156C}
{Cordes}, J.~M., \& {Lazio}, T.~J.~W. 2002, arXiv e-prints, astro.
\newblock \doarXiv{astro-ph/0207156}

\bibitem[{{Cordes} \& {Rickett}(1998)}]{1998ApJ...507..846C}
{Cordes}, J.~M., \& {Rickett}, B.~J. 1998, ApJ, 507, 846,
  \dodoi{10.1086/306358}

\bibitem[{{Cordes} {et~al.}(1985){Cordes}, {Weisberg}, \&
  {Boriakoff}}]{1985ApJ...288..221C}
{Cordes}, J.~M., {Weisberg}, J.~M., \& {Boriakoff}, V. 1985, ApJ, 288, 221,
  \dodoi{10.1086/162784}

\bibitem[{{Cordes} {et~al.}(1991){Cordes}, {Weisberg}, {Frail}, {Spangler}, \&
  {Ryan}}]{1991Natur.354..121C}
{Cordes}, J.~M., {Weisberg}, J.~M., {Frail}, D.~A., {Spangler}, S.~R., \&
  {Ryan}, M. 1991, Nature, 354, 121, \dodoi{10.1038/354121a0}

\bibitem[{{Cordes} {et~al.}(2016){Cordes}, {Wharton}, {Spitler}, {Chatterjee},
  \& {Wasserman}}]{2016arXiv160505890C}
{Cordes}, J.~M., {Wharton}, R.~S., {Spitler}, L.~G., {Chatterjee}, S., \&
  {Wasserman}, I. 2016, arXiv e-prints, arXiv:1605.05890.
\newblock \doarXiv{1605.05890}

\bibitem[{{Deller} {et~al.}(2008){Deller}, {Verbiest}, {Tingay}, \&
  {Bailes}}]{2008ApJ...685L..67D}
{Deller}, A.~T., {Verbiest}, J.~P.~W., {Tingay}, S.~J., \& {Bailes}, M. 2008,
  ApJL, 685, L67, \dodoi{10.1086/592401}

\bibitem[{{Deller} {et~al.}(2012){Deller}, {Archibald}, {Brisken},
  {Chatterjee}, {Janssen}, {Kaspi}, {Lorimer}, {Lyne}, {McLaughlin}, {Ransom},
  {Stairs}, \& {Stappers}}]{2012ApJ...756L..25D}
{Deller}, A.~T., {Archibald}, A.~M., {Brisken}, W.~F., {et~al.} 2012, ApJL,
  756, L25, \dodoi{10.1088/2041-8205/756/2/L25}

\bibitem[{{Deller} {et~al.}(2019){Deller}, {Goss}, {Brisken}, {Chatterjee},
  {Cordes}, {Janssen}, {Kovalev}, {Lazio}, {Petrov}, {Stappers}, \&
  {Lyne}}]{2019ApJ...875..100D}
{Deller}, A.~T., {Goss}, W.~M., {Brisken}, W.~F., {et~al.} 2019, ApJ, 875, 100,
  \dodoi{10.3847/1538-4357/ab11c7}

\bibitem[{{Dennison} {et~al.}(1999){Dennison}, {Simonetti}, \&
  {Topasna}}]{1999AAS...195.5309D}
{Dennison}, B., {Simonetti}, J.~H., \& {Topasna}, G.~A. 1999, in American
  Astronomical Society Meeting Abstracts, Vol. 195, American Astronomical
  Society Meeting Abstracts, 53.09

\bibitem[{{Desvignes} {et~al.}(2016){Desvignes}, {Caballero}, {Lentati},
  {Verbiest}, {Champion}, {Stappers}, {Janssen}, {Lazarus}, {Os{\l}owski},
  {Babak}, {Bassa}, {Brem}, {Burgay}, {Cognard}, {Gair}, {Graikou},
  {Guillemot}, {Hessels}, {Jessner}, {Jordan}, {Karuppusamy}, {Kramer},
  {Lassus}, {Lazaridis}, {Lee}, {Liu}, {Lyne}, {McKee}, {Mingarelli},
  {Perrodin}, {Petiteau}, {Possenti}, {Purver}, {Rosado}, {Sanidas}, {Sesana},
  {Shaifullah}, {Smits}, {Taylor}, {Theureau}, {Tiburzi}, {van Haasteren}, \&
  {Vecchio}}]{2016MNRAS.458.3341D}
{Desvignes}, G., {Caballero}, R.~N., {Lentati}, L., {et~al.} 2016, MNRAS, 458,
  3341, \dodoi{10.1093/mnras/stw483}

\bibitem[{{Finkbeiner}(2003)}]{2003ApJS..146..407F}
{Finkbeiner}, D.~P. 2003, ApJS, 146, 407, \dodoi{10.1086/374411}

\bibitem[{{Fonseca} {et~al.}(2014){Fonseca}, {Stairs}, \&
  {Thorsett}}]{2014ApJ...787...82F}
{Fonseca}, E., {Stairs}, I.~H., \& {Thorsett}, S.~E. 2014, ApJ, 787, 82,
  \dodoi{10.1088/0004-637X/787/1/82}

\bibitem[{{Gaensler} {et~al.}(2002){Gaensler}, {Jones}, \&
  {Stappers}}]{2002ApJ...580L.137G}
{Gaensler}, B.~M., {Jones}, D.~H., \& {Stappers}, B.~W. 2002, ApJL, 580, L137,
  \dodoi{10.1086/345750}

\bibitem[{{Gaensler} {et~al.}(2008){Gaensler}, {Madsen}, {Chatterjee}, \&
  {Mao}}]{2008PASA...25..184G}
{Gaensler}, B.~M., {Madsen}, G.~J., {Chatterjee}, S., \& {Mao}, S.~A. 2008,
  Publications of the Astronomical Society of Australia, 25, 184,
  \dodoi{10.1071/AS08004}

\bibitem[{{Gaustad} {et~al.}(2001){Gaustad}, {McCullough}, {Rosing}, \& {Van
  Buren}}]{2001PASP..113.1326G}
{Gaustad}, J.~E., {McCullough}, P.~R., {Rosing}, W., \& {Van Buren}, D. 2001,
  PASP, 113, 1326, \dodoi{10.1086/323969}

\bibitem[{{Guillemot} {et~al.}(2016){Guillemot}, {Smith}, {Laffon}, {Janssen},
  {Cognard}, {Theureau}, {Desvignes}, {Ferrara}, \& {Ray}}]{2016AA...587A.109G}
{Guillemot}, L., {Smith}, D.~A., {Laffon}, H., {et~al.} 2016, AAP, 587, A109,
  \dodoi{10.1051/0004-6361/201527847}

\bibitem[{{Haffner} {et~al.}(2003){Haffner}, {Reynolds}, {Tufte}, {Madsen},
  {Jaehnig}, \& {Percival}}]{2003ApJS..149..405H}
{Haffner}, L.~M., {Reynolds}, R.~J., {Tufte}, S.~L., {et~al.} 2003, ApJS, 149,
  405, \dodoi{10.1086/378850}

\bibitem[{{Harvey-Smith} {et~al.}(2011){Harvey-Smith}, {Madsen}, \&
  {Gaensler}}]{2011ApJ...736...83H}
{Harvey-Smith}, L., {Madsen}, G.~J., \& {Gaensler}, B.~M. 2011, \apj, 736, 83,
  \dodoi{10.1088/0004-637X/736/2/83}

\bibitem[{{Heiles}(1998)}]{1998ApJ...498..689H}
{Heiles}, C. 1998, ApJ, 498, 689, \dodoi{10.1086/305574}

\bibitem[{{Hill} {et~al.}(2005){Hill}, {Stinebring}, {Asplund}, {Berwick},
  {Everett}, \& {Hinkel}}]{2005ApJ...619L.171H}
{Hill}, A.~S., {Stinebring}, D.~R., {Asplund}, C.~T., {et~al.} 2005, ApJL, 619,
  L171, \dodoi{10.1086/428347}

\bibitem[{{Jennings} {et~al.}(2018){Jennings}, {Kaplan}, {Chatterjee},
  {Cordes}, \& {Deller}}]{2018ApJ...864...26J}
{Jennings}, R.~J., {Kaplan}, D.~L., {Chatterjee}, S., {Cordes}, J.~M., \&
  {Deller}, A.~T. 2018, \apj, 864, 26, \dodoi{10.3847/1538-4357/aad084}

\bibitem[{{Keith} {et~al.}(2013){Keith}, {Coles}, {Shannon}, {Hobbs},
  {Manchester}, {Bailes}, {Bhat}, {Burke-Spolaor}, {Champion}, {Chaudhary},
  {Hotan}, {Khoo}, {Kocz}, {Os{\l}owski}, {Ravi}, {Reynolds}, {Sarkissian},
  {van Straten}, \& {Yardley}}]{2013MNRAS.429.2161K}
{Keith}, M.~J., {Coles}, W., {Shannon}, R.~M., {et~al.} 2013, MNRAS, 429, 2161,
  \dodoi{10.1093/mnras/sts486}

\bibitem[{{Krishnakumar} {et~al.}(2015){Krishnakumar}, {Mitra}, {Naidu},
  {Joshi}, \& {Manoharan}}]{2015ApJ...804...23K}
{Krishnakumar}, M.~A., {Mitra}, D., {Naidu}, A., {Joshi}, B.~C., \&
  {Manoharan}, P.~K. 2015, ApJ, 804, 23, \dodoi{10.1088/0004-637X/804/1/23}

\bibitem[{{Lallement} {et~al.}(2015){Lallement}, {Vergely}, {Puspitarini},
  {Snowden}, {Galeazzi}, \& {Koutroumpa}}]{2015MmSAI..86..626L}
{Lallement}, R., {Vergely}, J.~L., {Puspitarini}, L., {et~al.} 2015, Memorie
  della Societa Astronomica Italiana, 86, 626

\bibitem[{{Lee} \& {Lee}(2019)}]{2019NatAs...3..154L}
{Lee}, K.~H., \& {Lee}, L.~C. 2019, Nature Astronomy, 3, 154,
  \dodoi{10.1038/s41550-018-0650-6}

\bibitem[{{Lee} \& {Jokipii}(1975)}]{1975ApJ...202..439L}
{Lee}, L.~C., \& {Jokipii}, J.~R. 1975, ApJ, 202, 439, \dodoi{10.1086/153994}

\bibitem[{{Manchester} {et~al.}(2005){Manchester}, {Hobbs}, {Teoh}, \&
  {Hobbs}}]{2005AJ....129.1993M}
{Manchester}, R.~N., {Hobbs}, G.~B., {Teoh}, A., \& {Hobbs}, M. 2005, \aj, 129,
  1993, \dodoi{10.1086/428488}

\bibitem[{{Nordgren} {et~al.}(1992){Nordgren}, {Cordes}, \&
  {Terzian}}]{1992AJ....104.1465N}
{Nordgren}, T.~E., {Cordes}, J.~M., \& {Terzian}, Y. 1992, AJ, 104, 1465,
  \dodoi{10.1086/116331}

\bibitem[{{Petroff} {et~al.}(2019){Petroff}, {Hessels}, \&
  {Lorimer}}]{2019A&ARv..27....4P}
{Petroff}, E., {Hessels}, J.~W.~T., \& {Lorimer}, D.~R. 2019, AAPR, 27, 4,
  \dodoi{10.1007/s00159-019-0116-6}

\bibitem[{{Phillips} \& {Clegg}(1992)}]{1992Natur.360..137P}
{Phillips}, J.~A., \& {Clegg}, A.~W. 1992, Nature, 360, 137,
  \dodoi{10.1038/360137a0}

\bibitem[{{Puspitarini} {et~al.}(2014){Puspitarini}, {Lallement}, {Vergely}, \&
  {Snowden}}]{2014A&A...566A..13P}
{Puspitarini}, L., {Lallement}, R., {Vergely}, J.~L., \& {Snowden}, S.~L. 2014,
  AAP, 566, A13, \dodoi{10.1051/0004-6361/201322942}

\bibitem[{{Rangelov} {et~al.}(2016){Rangelov}, {Pavlov}, {Kargaltsev},
  {Durant}, {Bykov}, \& {Krassilchtchikov}}]{2016ApJ...831..129R}
{Rangelov}, B., {Pavlov}, G.~G., {Kargaltsev}, O., {et~al.} 2016, ApJ, 831,
  129, \dodoi{10.3847/0004-637X/831/2/129}

\bibitem[{{Rangelov} {et~al.}(2017){Rangelov}, {Pavlov}, {Kargaltsev},
  {Reisenegger}, {Guillot}, {van Kerkwijk}, \& {Reyes}}]{2017ApJ...835..264R}
---. 2017, ApJ, 835, 264, \dodoi{10.3847/1538-4357/835/2/264}

\bibitem[{{Reardon} {et~al.}(2016){Reardon}, {Hobbs}, {Coles}, {Levin},
  {Keith}, {Bailes}, {Bhat}, {Burke-Spolaor}, {Dai}, {Kerr}, {Lasky},
  {Manchester}, {Os{\l}owski}, {Ravi}, {Shannon}, {van Straten}, {Toomey},
  {Wang}, {Wen}, {You}, \& {Zhu}}]{2016MNRAS.455.1751R}
{Reardon}, D.~J., {Hobbs}, G., {Coles}, W., {et~al.} 2016, MNRAS, 455, 1751,
  \dodoi{10.1093/mnras/stv2395}

\bibitem[{{Reynolds}(1989)}]{1989ApJ...339L..29R}
{Reynolds}, R.~J. 1989, ApJ, 339, L29, \dodoi{10.1086/185412}

\bibitem[{{Rickett}(1990)}]{1990ARA&A..28..561R}
{Rickett}, B.~J. 1990, ARAA, 28, 561,
  \dodoi{10.1146/annurev.aa.28.090190.003021}

\bibitem[{{Rickett} {et~al.}(2000){Rickett}, {Coles}, \&
  {Markkanen}}]{2000ApJ...533..304R}
{Rickett}, B.~J., {Coles}, W.~A., \& {Markkanen}, J. 2000, ApJ, 533, 304,
  \dodoi{10.1086/308637}

\bibitem[{{Ridley} {et~al.}(2013){Ridley}, {Crawford}, {Lorimer}, {Bailey},
  {Madden}, {Anella}, \& {Chennamangalam}}]{2013MNRAS.433..138R}
{Ridley}, J.~P., {Crawford}, F., {Lorimer}, D.~R., {et~al.} 2013, MNRAS, 433,
  138, \dodoi{10.1093/mnras/stt709}

\bibitem[{{Savage} \& {Wakker}(2009)}]{2009ApJ...702.1472S}
{Savage}, B.~D., \& {Wakker}, B.~P. 2009, ApJ, 702, 1472,
  \dodoi{10.1088/0004-637X/702/2/1472}

\bibitem[{{Schlafly} \& {Finkbeiner}(2011)}]{2011ApJ...737..103S}
{Schlafly}, E.~F., \& {Finkbeiner}, D.~P. 2011, ApJ, 737, 103,
  \dodoi{10.1088/0004-637X/737/2/103}

\bibitem[{{Schnitzeler}(2012)}]{2012MNRAS.427..664S}
{Schnitzeler}, D.~H.~F.~M. 2012, MNRAS, 427, 664,
  \dodoi{10.1111/j.1365-2966.2012.21869.x}

\bibitem[{{Shapiro-Albert} {et~al.}(2020){Shapiro-Albert}, {McLaughlin}, {Lam},
  {Cordes}, \& {Swiggum}}]{2020ApJ...890..123S}
{Shapiro-Albert}, B.~J., {McLaughlin}, M.~A., {Lam}, M.~T., {Cordes}, J.~M., \&
  {Swiggum}, J.~K. 2020, ApJ, 890, 123, \dodoi{10.3847/1538-4357/ab65f8}

\bibitem[{{Sharpless}(1959)}]{1959ApJS....4..257S}
{Sharpless}, S. 1959, ApJS, 4, 257, \dodoi{10.1086/190049}

\bibitem[{{Sivan}(1974)}]{1974A&AS...16..163S}
{Sivan}, J.~P. 1974, AAPS, 16, 163

\bibitem[{{Stinebring} {et~al.}(2019){Stinebring}, {Rickett}, \& {Koch
  Ocker}}]{2019ApJ...870...82S}
{Stinebring}, D.~R., {Rickett}, B.~J., \& {Koch Ocker}, S. 2019, ApJ, 870, 82,
  \dodoi{10.3847/1538-4357/aaef80}

\bibitem[{{Taylor} \& {Cordes}(1993)}]{1993ApJ...411..674T}
{Taylor}, J.~H., \& {Cordes}, J.~M. 1993, ApJ, 411, 674, \dodoi{10.1086/172870}

\bibitem[{{Taylor} \& {Manchester}(1977)}]{1977ApJ...215..885T}
{Taylor}, J.~H., \& {Manchester}, R.~N. 1977, \apj, 215, 885,
  \dodoi{10.1086/155426}

\bibitem[{{Toscano} {et~al.}(1999){Toscano}, {Britton}, {Manchester}, {Bailes},
  {Sandhu}, {Kulkarni}, \& {Anderson}}]{1999ApJ...523L.171T}
{Toscano}, M., {Britton}, M.~C., {Manchester}, R.~N., {et~al.} 1999, ApJL, 523,
  L171, \dodoi{10.1086/312276}

\bibitem[{{Tsai} {et~al.}(2015){Tsai}, {Simonetti}, {Akukwe}, {Bear},
  {Cutchin}, {Dowell}, {Gough}, {Kanner}, {Kassim}, {Schinzel}, {Shawhan},
  {Taylor}, {Yancey}, {Quezada}, \& {Kavic}}]{2015AJ....149...65T}
{Tsai}, J.-W., {Simonetti}, J.~H., {Akukwe}, B., {et~al.} 2015, AJ, 149, 65,
  \dodoi{10.1088/0004-6256/149/2/65}

\bibitem[{{van Leeuwen}(2007)}]{2007A&A...474..653V}
{van Leeuwen}, F. 2007, \aap, 474, 653, \dodoi{10.1051/0004-6361:20078357}

\bibitem[{{Vedantham} {et~al.}(2017){Vedantham}, {Readhead}, {Hovatta},
  {Koopmans}, {Pearson}, {Blandford}, {Gurwell}, {L{\"a}hteenm{\"a}ki},
  {Max-Moerbeck}, {Pavlidou}, {Ravi}, {Reeves}, {Richards}, {Tornikoski}, \&
  {Zensus}}]{2017ApJ...845...90V}
{Vedantham}, H.~K., {Readhead}, A.~C.~S., {Hovatta}, T., {et~al.} 2017, ApJ,
  845, 90, \dodoi{10.3847/1538-4357/aa7741}

\bibitem[{{Verbiest} {et~al.}(2009){Verbiest}, {Bailes}, {Coles}, {Hobbs}, {van
  Straten}, {Champion}, {Jenet}, {Manchester}, {Bhat}, {Sarkissian}, {Yardley},
  {Burke-Spolaor}, {Hotan}, \& {You}}]{2009MNRAS.400..951V}
{Verbiest}, J.~P.~W., {Bailes}, M., {Coles}, W.~A., {et~al.} 2009, MNRAS, 400,
  951, \dodoi{10.1111/j.1365-2966.2009.15508.x}

\bibitem[{{Vivekanand} \& {Narayan}(1982)}]{1982JApA....3..399V}
{Vivekanand}, M., \& {Narayan}, R. 1982, Journal of Astrophysics and Astronomy,
  3, 399, \dodoi{10.1007/BF02714882}

\bibitem[{{Yao} {et~al.}(2017){Yao}, {Manchester}, \&
  {Wang}}]{2017ApJ...835...29Y}
{Yao}, J.~M., {Manchester}, R.~N., \& {Wang}, N. 2017, ApJ, 835, 29,
  \dodoi{10.3847/1538-4357/835/1/29}

\end{thebibliography}

\appendix

\section{Derivation of the Likelihood Function}\label{app:theory}
The model for $n_e(\vec{x})$, the electron density along a LoS $\vec{x}(s)$, consists of a local mean density $\bar{n}_{e,0}(z)$ that exponentially depends on height above the Galactic plane plus variations from a turbulent medium following a Kolmogorov spectrum $n_k(\vec{x})$:
\begin{equation}
    n_e(\vec{x}) = \bar{n}_{e,0}(z) + n_{k}(\vec{x}),
\end{equation}
where $\bar{n}_{e,0}(z) = n_0e^{-|z|/z_0}$. The dispersion measure $N_e$ is given by 
\begin{equation}
    N_e(\hat{n},D) = \int_0^D n_e(\vec{x}(s)){\rm{d}}s
\end{equation}
and
\begin{equation}
    \langle N_e(\hat{n},D) \rangle = \int_0^D \bar{n}_{e,0}(z){\rm{d}}s,
\end{equation}
since $\langle n_{k}(\vec{x})\rangle=0$. Here $\hat{n}$ is the unit vector pointing along the LoS, and $s$ is the integration variable along the LoS. Computing the variance of $N_e$ requires the second moment:
\begin{equation}
\begin{split}
    \langle N_e(\hat{n},D)^2 \rangle &= \iint_0^D {\rm{d}}s {\rm{d}}s' \langle n_e(\vec{x}(s)) n_e(\vec{x}(s'))\rangle\\
    &= \iint_0^D {\rm{d}}s {\rm{d}}s' \langle (\bar{n}_{e,0}(z) + n_{k}(\vec{x}))(\bar{n}_{e,0}(z') + n_{k}(\vec{x}'))\rangle\\
    &= \langle N_e(\hat{n},D) \rangle^2 + \iint_0^D {\rm{d}}s {\rm{d}}s' \langle n_{k}(\vec{x}) n_{k}(\vec{x}') \rangle.
\end{split}
\end{equation}
The variance of $N_e$ is then
\begin{equation}
    \sigma_{N_e}^2(\hat{n},D) = \langle N_e^2 \rangle - \langle N_e \rangle^2
    = \iint_0^D {\rm{d}}s {\rm{d}}s' \langle n_{k}(\vec{x}) n_{k}(\vec{x}') \rangle.
\end{equation}
\indent The Kolmogorov contribution to the variance is
\begin{equation}
    \iint_0^D {\rm{d}}s {\rm{d}}s' \langle n_{k}(\vec{x}) n_{k}(\vec{x}') \rangle = \iint_0^D {\rm{d}}s {\rm{d}}s' \int {\rm{d}}^3q  P_n(\vec{q}).
\end{equation}
Performing the standard change of variables $\bar{s} = (s+s')/2$ and $\delta s = s-s'$ transforms the double integral $\int_0^D \int_0^D ds ds'$ to $\int_{-D}^D d\delta s \int_{|\delta s|/2}^{D-|\delta s|/2} d\bar{s}$ \citep[e.g.,][]{1998ApJ...507..846C}. We write $\vec{q} = \vec{q}_z + \vec{q}_\perp$, and the integral becomes
\begin{equation}
\begin{split}
    \iint_0^D {\rm{d}}s {\rm{d}}s' \langle n_{k}(\vec{x}) n_{k}(\vec{x}') \rangle &= \int_0^D {\rm{d}}\bar{s} C_n^2(\bar{s}) \int_{q_0}^\infty {\rm{d}}\vec{q}_\perp q_\perp^{-11/3}\\
    &= {\rm{SM}} \int_{q_0}^\infty {\rm{d}}q_\perp q_\perp^{-8/3}\\
    &= \frac{3}{5}{\rm{SM}}q_0^{-5/3},
\end{split}
\end{equation}
where SM = $\int_0^D {\rm{d}}\bar{s} C_n^2(\bar{s})$ is the scattering measure. We have assumed that the inner scale of the wavenumber spectrum is much smaller than the outer scale, and that the spectral coefficient $C_n^2$ varies slowly over the path length to the pulsar. The variance is simply
\begin{equation}
\begin{split}
    \sigma_{N_e}^2(\hat{n},D) &= \frac{3}{5}{\rm{SM}}q_0^{-5/3}.
\end{split}
\end{equation}
SM is usually reported in units of kpc m$^{-20/3}$, and wavenumber $q$ in units of m$^{-1}$. The variance $\sigma_{N_e}^2(\hat{n},D)$ has dimensions of DM$^2$, as expected. \\ 
\indent The model for $n_e(\vec{x})$ has only two free parameters: $\Theta = (n_0, z_0)^T$. We calculate SM by assuming $C_n^2 = C_{n,0}e^{-2|z|/z_0}$, with $C_{n,0} = 10^{-3.5}$ kpc m$^{-20/3}$. With the variance we can construct the likelihood function $\Lagr$ for $\Theta$ given observations of DM and distance (parallax) for a sample of pulsars. It is
\begin{equation}
    \Lagr(\Theta|D_j) = \prod_j \bigg\{[2\pi \sigma_{N_e}^2(\hat{n}_j,D_j)]^{-1/2}e^{-[{\rm{DM}}_j - \langle N_e(\hat{n}_j,D_j) \rangle]^2/2\sigma_{N_e}^2(\hat{n}_j,D_j)}\bigg\},
\end{equation}
where DM$_j$ is the observed DM of the $j$th pulsar in the sample. If we assume that the parallax error follows a Gaussian PDF $f_{\pi}(\pi)$, then $f_D(D) = (1/D^2)f_{\pi}(1/D)$ and the  parallax PDF for the $j$th pulsar will be
\begin{equation}
    f_{\pi_j}(\pi_j) = \frac{1}{\sqrt{2\pi\sigma_{\pi_j}^2}}e^{-(\pi-\bar{\pi}_j)^2/2\sigma_{\pi_j}^2}.
\end{equation}
The resulting distance PDF will be
\begin{equation}
    f_{D_j}(D_j) = \frac{1}{D_j^2\sqrt{2\pi\sigma_{\pi_j}^2}}e^{-((1/D)-(1/\bar{D}_j))^2/2\sigma_{\pi_j}^2},
\end{equation}
where $\bar{\pi}_j$ is the measured parallax of the $j$th pulsar (in arcseconds) and $\bar{d}_j$ is the measured distance (in parsecs). While the parallax PDF is Gaussian, the distance PDF is not. The full likelihood function is
\begin{equation}\label{eq:likefunc}
    \Lagr(\Theta) = \prod_j \int f_{D_j}(D_j) {\rm{d}}(D_j) \bigg\{[2\pi \sigma_{N_e}^2(\hat{n}_j,D_j)]^{-1/2}e^{-[{\rm{DM}}_j - \langle N_e(\hat{n}_j,D_j) \rangle]^2/2\sigma_{N_e}^2(\hat{n}_j,D_j)}\bigg\}.
\end{equation}

\section{Pulsar Outliers and Other Notable Lines of Sight}\label{app:outliers}

Several pulsars in our sample have scattering structures intervening their lines of sight but have DMs consistent with the PPK model, implying that these scattering structures have minimal impact on the integrated column density. Two pulsars in the sample have small DM departures from the model that could be caused by discrete structures or turbulent density fluctuations. Six pulsars in the sample have significant DM excess from the model but do not have complementary information about their lines of sight. All of these cases are summarized below.

\indent \textbf{J0437$-$4715} has a small DM excess of $0.42 \pm 0.13$ pc cm$^{-3}$. \cite{2016ApJ...818...86B} observe scintillation arcs in the secondary spectrum of J0437$-$4715 at 192 MHz and 732 MHz, and estimate the scattering screen's location to be at 115 pc, consistent with the edge of the Local Bubble. This pulsar also has one of the first bow shocks detected in H$\alpha$ \citep{1995ApJ...440L..81B} and has also been seen in far ultraviolet \citep{2016ApJ...831..129R}. The transverse velocity of the pulsar is $v_\perp = 104.71 \pm 0.95$ km/s \citep{2008ApJ...685L..67D}, and the standoff angular scale is about $9.3^{\prime \prime}$ \citep{2014ApJ...784..154B}, which implies a standoff radius of about 0.007 pc ($\approx 1400$ AU) if the bow shock is seen face-on. The bow shock's upstream neutral hydrogen column density is $\approx 0.2$ cm$^{-3}$ \citep{2014ApJ...784..154B}. For this pulsar, we find a lower limit on the clump parameter $L_{\rm min} [gf/\zeta(1 + \epsilon^2)] > 0.05$ pc. Since $g$, $f$, and $\epsilon$ are $\leq 1$ and $\zeta \geq 1$, the minimum clump size is too large for the bow shock and/or other plasma screens responsible for the scintillation arcs to dominate the excess DM. The DM variance due to turbulent density fluctuations is $\sigma_{N_e}^2 = (0.95\hspace{0.05in} {\rm pc} \hspace{0.05in} {\rm cm}^{-3})^2$ for this LoS, so the DM excess for this pulsar could be the result of turbulence.

\indent \textbf{J0814+7429 (B0809+74)} does not exhibit significant departure in DM from the PPK model. \cite{2000ApJ...533..304R} demonstrate that weak interstellar scintillation detected from this pulsar is consistent with both an extended scattering medium along the LoS or a localized compact structure, but the former is favored due to the weak scattering observed, along with the high relative velocity that would be required for a compact scattering region. 

\indent \textbf{J0953+0755 (B0950+08)} has a small DM deficit from  the PPK model of DM$-$DM$_{\rm PPK} = -0.66 \pm 0.21$ pc cm$^{-3}$, indicating a void along the LoS. Its distance of 262 pc and latitude of $43.7^\circ$ mean that a significant portion of its LoS traverses the Local Bubble (LB), which is estimated to have a mean electron density of $\approx 0.005$ cm$^{-3}$, about an order of magnitude smaller than the WIM \citep[e.g.,][]{1998ApJ...500..262B}. Assuming the LB is about 200 pc long and the total DM is a combination of the LB medium and the PPK medium, then the LB should contribute about 2 pc cm$^{-3}$ to the pulsar's DM. Scintillation observations show that ${\rm log}(\cnsq) \sim -4.5$, an order of magnitude smaller than for other nearby pulsars \citep{1992Natur.360..137P, 1998ApJ...500..262B}. The cloudlet model relates SM to DM (c.f. Eq.~\ref{eq:SMDM}). The value of $\cnsq$ implies SM $\approx 8.2 \times 10^{-6}$ kpc m$^{-20/3}$, which combined with DM gives $F_c \approx 0.1$ for the whole LoS. Breaking SM and DM into separate components for the WIM and LB also yields $F_c \approx 0.1$ in both components, which is higher than the value used for the LB in NE2001, $F_c = 0.01$. It is somewhat surprising that the DM deficit from the PPK model along this LoS is so small, not only because of the small SM but also because scatter-broadening of giant pulses from B0950+08 are consistent with a power spectral index $\alpha \approx 0.4$, a significant deviation from a Kolmogorov spectrum \citep{2015AJ....149...65T}.

\indent \textbf{J1136+1551 (B1133+16)} has a DM consistent with the PPK model. Scintillation from this pulsar has been extensively observed and is broadly consistent with a scattering screen at the edge of the Local Bubble \citep{2019ApJ...870...82S}. 

\indent \textbf{J1543-0620 (B1540$-$06)} exhibits a DM deficit from the PPK model DM$-$DM$_{\rm PPK} = -8.4 \pm 2.6$ pc cm$^{-3}$. No discrete structures were found along this LoS in catalogs of H$\alpha$, molecular line, or radio surveys.

\indent \textbf{J1840+5640 (B1839+56)} has a DM excess from the PPK model DM$-$DM$_{\rm PPK} = 8.4 \pm 1.3$ pc cm$^{-3}$. Evidence for scattering along this LoS has not been found. No discrete structures were found along this LoS in catalogs of H$\alpha$, molecular line, or radio surveys. The measured $I_{\rm{H}\alpha}$ for this LoS gives a characteristic clump size $L_{\max} [gf/\zeta(1 + \epsilon^2)] = 14 \pm 4$ pc. 

\indent \textbf{J2006$-$0807 (B2003$-$08)} has a DM excess DM$-$DM$_{\rm PPK} = 5.3 \pm 2.1$ pc cm$^{-3}$. The H$\alpha$ intensity for this LoS gives a characteristic clump size $L_{\max} [gf/\zeta(1 + \epsilon^2)] = 4 \pm 2$ pc. Evidence for scattering along this LoS has not been found. No discrete structures were found along this LoS in catalogs of H$\alpha$, molecular line, or radio surveys.

\indent \textbf{J2124$-$3358} also has a DM consistent with the PPK model. This pulsar has a bow shock detected in both H$\alpha$ \citep{2002ApJ...580L.137G} and far ultraviolet \citep{2017ApJ...835..264R}, and is traveling through a medium with mean density $n_e \approx 0.8$-1.3 cm$^{-3}$ and a density gradient of up to 10 cm$^{-3}$ over less than 0.02 pc \citep{2002ApJ...580L.137G}. 

\indent \textbf{J2129$-$5721} has a DM excess from the PPK model DM$-$DM$_{\rm PPK} = 10 \pm 2$ pc cm$^{-3}$. Evidence for scattering along this LoS has not been found. No discrete structures were found along this LoS in catalogs of H$\alpha$, molecular line, or radio surveys. The upper limit on H$\alpha$ intensity for this LoS gives a characteristic clump size $L_{\max} [gf/\zeta(1 + \epsilon^2)] > 22$ pc.

\indent \textbf{J2144$-$3933} has a DM excess from the PPK model DM$-$DM$_{\rm PPK} = 1.0 \pm 0.1$ pc cm$^{-3}$. Evidence for scattering along this LoS has not been found. No discrete structures were found along this LoS in catalogs of H$\alpha$, molecular line, or radio surveys. The upper limit on $I_{\rm{H}\alpha}$ for this LoS yields a characteristic clump size $L_{\min} [gf/\zeta(1 + \epsilon^2)]>0.3$ pc.

\indent \textbf{J2305+3100 (B2303+30)} has a DM excess DM$-$DM$_{\rm PPK} = 12 \pm 4$ pc cm$^{-3}$. The measured H$\alpha$ intensity for this LoS gives characteristic clump size $L_{\max} [gf/\zeta(1 + \epsilon^2)] = 29 \pm 13$ pc. Evidence for scattering along this LoS has not been found. No discrete structures were found along this LoS in catalogs of H$\alpha$, molecular line, or radio surveys.

\listofchanges

\end{document}